\documentclass[12pt,prd]{revtex4}

\usepackage{graphicx}
\usepackage{amsmath}

\begin{document}
%\preprint{RBRC-1005} 

\title{Phase structure and Hosotani mechanism
\\in gauge theories with compact dimensions revisited}

\author{Kouji Kashiwa} 
\email{kashiwa@ribf.riken.jp}
\affiliation{RIKEN/BNL, Brookhaven National Laboratory, Upton, NY 11973}

\author{Tatsuhiro Misumi} 
\email{tmisumi@bnl.gov} 
\affiliation{Department of Physics, Brookhaven National Laboratory, Upton, NY 11973}

\begin{abstract} 
We investigate the phase structure of $SU(3)$ gauge theory
in four and five dimensions with one compact dimension by using
perturbative one-loop and PNJL-model-based effective potentials, 
with emphasis on spontaneous gauge symmetry breaking. 
When adjoint matter with the periodic boundary condition is introduced,
we have rich phase structure in the quark-mass and compact-size space
with gauge-symmetry-broken phases, called the $SU(2)\times U(1)$ split
and the $U(1)\times U(1)$ re-confined phases.
Our result is qualitatively consistent with the recent lattice calculations. 
When fundamental quarks are introduced in addition to adjoint quarks, 
the split phase becomes more dominant and larger as a result of explicit center symmetry breaking. 
We also show that another $U(1)\times U(1)$ phase 
(pseudo-reconfined phase) with negative vacuum expectation 
value of Polyakov loop exists in this case. 
We study chiral properties in these theories and show 
that chiral condensate gradually decreases and chiral symmetry is slowly restored 
as the size of the compact dimension is decreased. 
\end{abstract}

\maketitle

\newpage

%%%%%%%%%Introduction%%%%%%%%%%%%%%%%%%%

\section{Introduction}
\label{sec:intro}

Now that the Higgs-like particle has been discovered in Large Hadron Collider (LHC) \cite{AT1, CMS1},
one of primary interests of particle physics is to understand mechanism of
dynamical electroweak symmetry breaking. To settle down the unsolved issues including
the hierarchy problem, there have been proposed several promising courses 
including supersymmetric, composite-Higgs and extra-dimensional models. 
Among them, the gauge-Higgs unification with extra dimensions \cite{Man1, H1} produces
brilliant dynamics of gauge symmetry breaking, called the Hosotani mechanism 
\cite{H1,H2, DMc1, HIL1, Hat1, H3, H4}, with the Higgs particle
as an extra-dimensional component of the gauge field:
when adjoint fermions are introduced with periodic boundary condition (PBC)
in gauge theory with a compact dimension, the compact-space component 
of the gauge field can develop a non-zero vacuum expectation value (VEV), 
which breaks the gauge symmetry to its subgroup spontaneously.
This phenomenon originates in the non-abelian Aharonov-Bohm effect.
There have been proposed a number of BSM models directly and indirectly based on this mechanism
\cite{ACP1, MT1, HMTY1, ST1, FOOS1, ST2, HKT1, IMST1, HNU1, HTU1, HTU2, HH1, IKY0}

Recently, the same phenomenon has been observed in a different context. 
As well-known, the imaginary time dimension is compacted in the
finite-temperature (Quantum Chromodynamics) QCD.
When the adjoint fermion is introduced with PBC in the theory, 
it was shown that exotic phases appear \cite{U1, U2, O1, MO1, MO2, MO3, NO1, CD1}, 
where the color trace of Polyakov line takes non-trivial VEV 
leading to the dynamical gauge symmetry breaking.
This can be interpreted as $R^{3}\times S^{1}$ realization of 
the Hosotani mechanism. Further study is now required to understand detailed
properties and phase diagram for this case.
As for the five-dimensional gauge theory on $R^{4}\times S^{1}$, 
lattice simulation is still at the stage to understand the nature of 5D pure Yang-Mills theory.
First attempt was done in Ref.~\cite{C1} 
and recent progress was shown in Ref.~\cite{KNO1, JN1, BBB1, LPS1, CW1, BCW1, EKM1, EFK1, DFKK1, DFKKN1, DFK1, FFKLV1, IK1, IK2, IK3, IKL1, KIL1, IK4, IK5, DKR1, dFKP1, KLR1, Del1, IKY1}.
At the present, we have no detailed investigation on the phase structure of
the five-dimensional gauge theory with dynamical quarks.

The purpose of our work is to understand properties of gauge theories with quarks on 
$R^{3}\times S^{1}$ and $R^{4}\times S^{1}$
by using effective theories and show a guideline for further lattice simulations.
We focus on the phase structure in $SU(3)$ gauge theory 
by mainly using the perturbative one-loop effective potential.
When we look into chiral properties, the Polyakov-loop-extended 
Nambu--Jona-Lasinio (PNJL) model \cite{GO1,Fuku1, K1, HK1} is introduced.
We obtain the phase diagram in the quark-mass and compacted-size space
for several different choices of fermion representations and boundary conditions.
In the theory with adjoint fermions with PBC,
we find rich phase diagram with unusual phases 
where $SU(3)$ gauge symmetry is spontaneously broken
to $SU(2)\times U(1)$ (split phase) or $U(1)\times U(1)$ (re-confined phase).
This result is consistent with the recent lattice simulations \cite{MO1, CD1} 
and the perturbative calculations for massless quarks \cite{H3, H4}.
We show that chiral condensate is gradually decreased and 
the chiral symmetry is slowly restored with the
compacted size being decreased although small chiral transitions
coincide with the deconfined/split and split/reconfined transitions.
We also study phase diagram for the case with both fundamental and adjoint quarks,
and show that the split phase becomes more dominant as a result of the explicit breaking
of $Z_{3}$ center symmetry. 
This result indicates that the $SU(3)\to SU(2)\times U(1)$ phase,
which is of significance in terms of phenomenology,  can be controlled
by adjusting the number of fundamental flavors.  
In this case we point out that another unusual ($U(1)\times U(1)$) phase 
with negative VEV of color-traced Polyakov loop emerges, 
which we call ``pseudo-reconfined phase".

We note similar investigation on the phase diagram of compact-space 
gauge theory with PBC fermions were shown in \cite{U1, U2, O1, MO2, NO1}, 
where the authors focus on the volume independence of the vacuum
structure from the viewpoint of the large N reduction
\cite{EK1,  KUY1, U1, U2, UY1, BS1, B1, BS2, PU1}.  
In order to reproduce the confined phase at low temperature,
they used the gluon effective potentials with the mass-dimension parameter, 
which leads to explicit gauge symmetry breaking.
Since these effective potentials do not suit our purpose of studying 
spontaneous gauge symmetry breaking due to the Hosotani mechanism,
we mainly use setups with manifest gauge symmetry.
It is also notable that gauge symmetry breaking for all clasical Lie groups with PBC matters
are classified from topological viewpoints in \cite{AU1}.

This paper is organized as following.
In Sec.~\ref{sec:EP}, we show our setups including the one-loop
potential and the PNJL model.
Sec.~\ref{sec:PS4} shows our numerical results for four dimensional cases.
Sec.~\ref{sec:PS5} shows the results for five-dimensional cases. 
In Sec.~\ref{sec:OB} we discuss how our predictions can be checked in the lattice simulations. 
Sec.~\ref{sec:sum} is devoted to summary.

%%%%%%%%%%%EP%%%%%%%%%%%%%%%%%%%%%

\section{Effective potential}
\label{sec:EP}
In this section we discuss our setups for SU(N) gauge theory.
The one-loop effective potential is composed of two parts, gluonic and quark contributions.
Since our goal is to study the phase diagram in terms of
Wilson-line phases (Polyakov-line phases), $q_{i}$ $(i=1,2,...,N)$,
we will write the effective potential as a function of  these variables.
Besides the one-loop potential, we consider the contribution from 
the chiral part, which originates in the NJL-type four-point interactions.
The total effective potential is then regarded as that of 
the PNJL model \cite{Fuku1} as a function of the $q_{i}$ and the chiral condensate $\sigma$.
Depending on our purposes, we in some cases use only the perturbative
one-loop effective potential,
and in other cases use the PNJL effective potential. 
We also comment on possible deformation reflecting non-perturbative effects 
in the gluonic contribution.

\subsection{SU(N) in four dimensions}
\label{subsec:D4}

We begin with one-loop effective potential for $SU(N)$ gauge theory.
In four dimensions, the finite-temperature one-loop effective potential 
for gauge bosons and fermions is well studied in Ref.~\cite{GPY1,W1}.
What we do in the present study is just to regard the compacted 
imaginary-time direction as one of the spatial direction, and
to replace $T$ by $1/L$ where $L$ is a size of the compacted
dimension.

Firstly, we rewrite the $SU(N)$ gauge boson field as
\begin{align}
A_\mu &= \langle A_y \rangle + \tilde{A}_\mu,
\end{align}
where $y$ stands for a compact direction and $\langle A_y \rangle$ 
is a vacuum expectation value (VEV) while $\tilde{A}_\mu$ is fluctuation from it.
The VEV can be replaced by
\begin{align}
\langle A_y \rangle &=\frac{2 \pi}{gL} q,
\end{align}
where $q$'s color structure is $\mathrm{diag}(q_1,q_2,...,q_{N})$ with 
$q_1+q_2+\cdot\cdot\cdot+q_{N-1}+q_{N}=0$.
We note that eigenvalues of $q_{i}$ 
are invariant under all gauge transformations preserving boundary conditions.
Then we can easily observe spontaneous gauge symmetry breaking from 
values of $q_{i}$ in the vacuum.
For detailed argument on gauge transformation for this topic, see \cite{H3} for example.
The gluon one-loop effective potential ${\cal V}_g$ is expressed as
\begin{align}
{\cal V}_g
&= - \frac{2}{L^4 \pi^2} \sum_{i,j=1}^N \sum_{n=1}^{\infty} 
     \Bigl( 1 - \frac{1}{N} \delta_{ij} \Bigr)
     \frac{\cos( 2 n \pi q^{ij})}{n^4} 
\end{align}
where $q_{ij} = ( q_i - q_j )$
and $N$ is the number of color degrees of freedom.

The contribution from massless fundamental quarks
${\cal V}_f$ is given by
\begin{align}
{\cal V}_f
&= \frac{4 N_f}{L^4 \pi^2} \sum_{i=1}^N \sum_{n=1}^\infty
   \frac{\cos[2 \pi n (q_i + 1/2)]}{n^4},
\end{align}
where $N_f$ is the number of fundamental flavors.
Depending on boundary conditions, we should replace 
$q_i + 1/2$ by $q_i + \phi$.
For example, the choice of $\phi=0$ describes
quarks with periodic boundary conditions.
From here, we denote ${\cal V}_f^\phi$ as 
effective potential of the fundamental fermion with boundary angle $\phi$.
The contribution from massive fundamental quarks
is expressed by using the second kind of the modified Bessel function $K_2(x)$ as
\begin{align}
{\cal V}_f^\phi(N_{f},m_f) &=
\frac{ 2 N_f m_{f}^{2}}{\pi^2L^2} \sum_{i=1}^N \sum_{n=1}^\infty 
\frac{K_2 ( n m_{f} L )}{n^2}
\cos [2 \pi n (q_i + \phi)],
\end{align}
where $m_{f}$ is the fundamental fermion mass. (We assume the same mass for all flavors.)
Here the second kind of the modified Bessel function $K_\nu(x)$ is defined as
\begin{align}
K_\nu (x) 
&= \frac{\sqrt{\pi} (x/2)^\nu}{\Gamma(\nu+1/2)}
   \int_1^\infty e^{-xt} (t^2-1)^{\nu-\frac{1}{2}} dt,
\end{align}
where $\Gamma(x)$ is the gamma function.
The adjoint quark contribution ${\cal V}_a^\phi$ 
can be easily obtained by the following replacement,
\begin{align}
{\cal V}_a^\phi (N_{a}, m_{a}) &=
\frac{ 2 N_a m_{a}^{2}}{\pi^2L^2} \sum_{i,j=1}^N \sum_{n=1}^\infty 
\Bigl(  1 - \frac{1}{N} \delta_{ij} \Bigr)
\frac{K_2 ( n m_{a} L )}{n^2}
\cos [2 \pi n (q_{ij} + \phi)],
\label{Re}
\end{align}
where $N_a$ and $m_{a}$ are the number of flavors and the mass for adjoint fermions.
For the gauge theory with $N_{f}$ fundamental and $N_{a}$ adjoint fermions with
arbitrary boundary conditions,
the total one-loop effective potential is given by
\begin{equation}
{\cal V} = {\cal V}_{g}+{\cal V}_{f}^{\phi}(N_{f}, m_f)+{\cal V}_a^{\phi}(N_{a}, m_a).
\label{eq_ep_pert}
\end{equation}
This total one-loop effective potential contains eight parameters including
the compactification scale $L$, the number of colors $N$, the fermion masses $m_{f}$, $m_{a}$, the number of flavors $N_{f}$, $N_{a}$, and the boundary conditions $\phi$ for two kinds of matter fields.
All through the present study we keep $N=3$, then obtain phase diagram in $1/L$-$m_{a}$ space
with $m_{f}$, $N_{f}$, $N_{a}$ and $\phi$ fixed to several values.
The reason we change $m_{a}$ while fixing $m_{f}$ is that gauge symmetry phase diagram is more sensitive to the former than the latter.
Up to this point, we work in first-principle perturbative calculations and have no 
model-parameter-fixing process dictated by physical results or numerical calculations.

We here comment on ability and limitation of the perturbative one-loop effective potential.
This weak-coupling-limit potential at least works to investigate weak-coupling physics.
In this study a small compactification scale $L$ (or high temperature $T\sim 1/L$) corresponds 
to weak coupling, and the $1/L$-$m_{a}$ phase diagram obtained from the effective potential 
is valid at smaller $L$ while the strong-coupling physics near larger $L$ or small temperature 
cannot be reproduced.
We note that gauge symmetry breaking due to Hosotani mechanism takes place 
even at weak-coupling due to the compact dimension topology \cite{H1,H2, H3, H4}.
Indeed, the exotic phases in the recent lattice simulations \cite{CD1} have been shown to be
at weak-coupling regime. Our potential thus work to reveal properties of these exotic phases.

Now we consider contribution from the chiral sector.
We introduce the four-point interaction at the action level as
\begin{equation}
g_{S}[(\bar{\psi} \psi)^{2}
     +(\bar{\psi} i \gamma_{5} {\vec \tau} \psi)^{2}],
\label{gs}
\end{equation}
where $\psi$ is a two-flavor fermion field and $g_{S}$ is the effective coupling constant
with the mass dimension minus two. (We here concentrate on the two-flavor case.)
The introduction of the above term (\ref{gs}) leads to 
addition of the following zero-temperature contribution 
to the one-loop effective potential (\ref{eq_ep_pert}),
\begin{equation}
{\cal V}_{\chi} = g_{S} \sigma^{2}-{d_{R}\Lambda^{4}\over{4\pi^{2}}}\left[
\left(2+{m_{c}^{2}\over{\Lambda^2}}\right)\sqrt{1+{m_{c}^{2}\over{\Lambda^2}}}+{m_{c}^{4}\over{\Lambda^4}}
\log\left({m_{c}/\Lambda\over{1+\sqrt{1+m_{c}^2 /\Lambda^{2}}}}\right)
\right],
\end{equation}
where $\Lambda$ is a cutoff scale of the effective theory and 
$m_{c}$ stands for constituent quark mass $m_{c}=m-2g_{S}\sigma$
with $\sigma$ being chiral condensate \cite{NO1}.
Depending on fundamental or adjoint representations, $d_{R}$ takes $N$ or $N^2 -1$
respectively. In the case with both fundamental and adjoint quarks,
we need two different sets of chiral sectors. 
The total effective potential combined with the chiral contribution is given by
\begin{equation}
{\cal V}_\mathrm{total} = {\cal V} + {\cal V}_{\chi}.
\label{PNJL-like}
\end{equation}
We note that this total effective potential is identical to that of the PNJL model \cite{Fuku1}
adopting the one-loop potential as the gluon contribution,
\begin{equation}
{\cal L}_{\rm PNJL}= \bar{\psi}(\gamma_{\mu}D_{\mu}+m)\psi
-g_{S}[(\bar{\psi} \psi)^{2}+(\bar{\psi} i\gamma_{5}{\vec \tau}\psi)^{2}]
+{\cal V}_{g},
\end{equation}
with $D_{j}=\partial_{j}, \, D_{4}=\partial_{4}+iA_{4}$.
We thus call the total effective potential (\ref{PNJL-like}) PNJL or PNJL-based effective potential.

In the present study we will use this PNJL-based potential to study gauge theory with adjoint or fundamental quarks with periodic boundary conditions(PBC).
In this model, we have two model parameters, the cutoff scale $\Lambda$ 
and the effective-coupling $g_{S}$ in addition to the eight parameters 
in the one-loop perturbative potential shown below (\ref{eq_ep_pert}).
These two parameters are the very model parameters and should be fixed
to reproduce physical results or first-principle lattice calculations.
On the other hand, the lattice QCD simulation on the compactified space has been done
as finite-temperature lattice QCD, but for only anti-periodic boundary conditions 
(aPBC) for quarks.
Thus, we choose the following criterion for fixing the two parameters for PBC quarks;
we first fix them so as to reproduce the zero-temperature constituent mass
and the chiral critical temperature in the lattice finite-temperature 
QCD with aPBC matters \cite{KL1}, then we use the same parameter set for the PBC case.
According to \cite{KL1}, the zero-temperature constituent mass for aPBC adj. QCD
is given by
\begin{equation}
m_{c}(T=0, m_{a}=0)=2.322{\rm GeV},
\end{equation}
and the chiral phase transition takes place at
\begin{equation}
1/L_{CT}=T_{CT}\sim 2 {\rm GeV},
\end{equation}
where $L_{CT}$ and $T_{CT}$ are critical values for chiral transition.
We can fix $\Lambda$ and $g_{S}$ by using these lattice results.
This method is also applied to the fundamental quarks.

Apart from the chiral contribution, the PNJL potential we obtained is based on 
perturbative calculations, and chiral properties we will obtain from it is again valid 
for a small $L$ region.
In addition to this limitation, we have the cutoff scale $\Lambda$ in this model.
It means that we cannot apply the model beyond this cutoff. 
The model is thus valid for intermediate compactification scale as $0\ll 1/L <\Lambda$.
In the next section we will find specific chiral properties within this limitation. 

We here comment on non-perturbative modification of the PNJL model.
In the standard use of the PNJL model, the gluonic contribution ${\cal V}_{g}$
is replaced by the ``non-perturbatively"-deformed ones:
in order to mimic the confinement/deconfinement 
phase transition in the study on QCD phase diagram,, 
the one-loop gluon potential should be replaced by some nonperturbative versions.
We have several schemes including the simple scale introduction in Ref.~\cite{NO1}, 
center-stabilized potential in \cite{MO3, MeMO1, MeO1},
the one-loop {\it ansatz} in Ref.~\cite{Rob1} 
and the strong-coupling lattice potential in Ref.~\cite{Fuku1, RRTW1}.
The first one in \cite{NO1} is given by
\begin{align}
{\cal V}_{g}^\mathrm{np1}
&=- \frac{2}{L^4 \pi^2} \sum_{i,j=1}^N \sum_{n=1}^{\infty} \Bigl( 1 - \frac{1}{N} \delta_{ij} \Bigr)
     \frac{\cos( 2 n \pi q^{ij})}{n^4} \,+\, \frac{M^2}{2\pi^2L^2} \sum_{i,j=1}^N \sum_{n=1}^\infty
  \Bigl( 1 - \frac{1}{N} \delta_{ij} \Bigr) \frac{\cos(2n\pi q^{ij})}{n^2},
\label{g_np}
\end{align}
where the second term is the modification incorporating non-perturbative effect 
and the mass scale $M$ works as a scale for confinement/deconfinement transtion.
The center-stabilized potential in \cite{UY1, MO3, MeMO1, MeO1} has a similar form given as
\begin{align}
{\cal V}_{g}^\mathrm{np2}
&=- \frac{2}{L^4 \pi^2} \sum_{i,j=1}^N \sum_{n=1}^{\infty} \Bigl( 1 - \frac{1}{N} \delta_{ij} \Bigr)
     \frac{\cos( 2 n \pi q^{ij})}{n^4} \,+\,{1\over{L}}\sum^{[N/2]}_{n=1}
    \sum_{i,j=1}^N a_{n} \Bigl( 1 - \frac{1}{N} \delta_{ij} \Bigr)\cos( 2 n \pi q^{ij}),
\label{g_np2}
\end{align}
where $a_{n}$ in the deformation term is a mass dimension-3 model parameter, which works as the phase transition scale in this case.
The one-loop ansatz potential is based on simple introduction of the scale $M$ to the original gluon potential as
\begin{align}
{\cal V}_{g}^\mathrm{np3}
&=- \frac{2M^2}{L^2 \pi^2} \sum_{i,j=1}^N \sum_{n=1}^{\infty} \Bigl( 1 - \frac{1}{N} \delta_{ij} \Bigr)
     \frac{\cos( 2 n \pi q^{ij})}{n^4},
\label{g_np3}
\end{align}
where some of compactification scale $L$ is just replaced by $M$ in the original potential.
The potential from the lattice strong-coupling expansion is non-perturbative deformation adopted
in the original PNJL model \cite{Fuku1}, which is given by
\begin{align}
{\cal V}_{g}^\mathrm{np4} = -{4\over{a^{3}L}}e^{-\sigma aL}({\rm Tr}P)^{2}-
{1\over{a^{3}L}}\log\Big[ -({\rm Tr}P)^{4}+8{\rm Re}({\rm Tr}P)^{3}-18({\rm Tr}P)^{2}+27 \Big],
\label{g_np4}
\end{align}
where $1/a$ stands for the cutoff scale and $\sigma$ for the string tension for confinement.
$P=\mathcal{P}\exp [ig\int dy A_{y}]\sim{\rm diag}[e^{2\pi i q_{1}},...,e^{2\pi i q_{N}}]$ is a compactified dimension Polyakov (Wilson) loop. 
These two model parameters work as the scale for the phase transition.
All the modifications able to reproduce the phase transition, however, it is notable that
the $SU(N)$ gauge symmetry is explicitly broken due to the mass-dimensionful 
model parameters in them.
It means that they do not suit our purpose of classifying the gauge-broken phases.
We also emphasize that, in the first place, the exotic phases including gauge-broken phases are likely to emerge at weak-coupling region \cite{CD1} and we do not need the above deformations for the strong-coupling region for our purpose.
Thus, we will mainly adopt one-loop effective potential as the gluonic contribution in our PNJL model
and just discuss how the phase diagram is changed by the deformation in the next section.

\subsection{SU(N) in five dimensions}
\label{subsec:D5}

In the five-dimensional case, most of the setup is parallel to the
four-dimensional case except for difference of mass dimensions of fields and parameters.
We here show the one-loop effective potential of gluon and quarks below.
The five-dimensional one-loop effective potential in the gluon
sector is given by
\begin{align}
{\cal V}_g
&= - \frac{9}{4\pi^2L^5} \sum_{i,j=1}^N \sum_{n=1}^{\infty} 
     \Bigl( 1 - \frac{1}{N} \delta_{ij} \Bigr)
     \frac{\cos( 2 \pi n q_{ij})}{n^5},
\end{align}
which has $1/L^5$ dimension.
The effective potential for massless fermions in this case is given by
\begin{align}
{\cal V}_f^{1/2}
&= -2 T \int \frac{dk}{ 8\pi^2} 
    k^3 \Bigl[ \ln \Bigl( 1 + e^{-L|{\bf k}| + 2 i \pi q^{ij}} \Bigr) 
             + \ln \Bigl( 1 + e^{-L|{\bf k}| - 2 i \pi q^{ij}} \Bigr) \Bigr]
\nonumber\\
&= \frac{3}{L^5 \pi^2} \sum_{n=1}^\infty
   \frac{\cos [2 \pi n (q^{ij}+1/2) ]}{n^5}.
\label{Integral}
\end{align}
The effective potential of the massless fundamental and adjoint fermions with arbitrary
boundary condition is expressed as
\begin{align}
{\cal V}_f^\phi 
&= \frac{3 N_f}{\pi^2L^5} \sum_{i=1}^N \sum_{n=1}^\infty 
   \frac{\cos[ 2 \pi n (q_i + \phi ) ]}{n^5},
\\
{\cal V}_a^\phi 
&= \frac{3 N_a}{\pi^2L^5} \sum_{i,j=1}^N \sum_{n=1}^\infty 
   \Bigl( 1 - \frac{1}{N} \delta_{ij} \Bigr)
   \frac{\cos[ 2 \pi n (q_{ij} + \phi ) ]}{n^5}.
\end{align}

For the massive fermion, we should replace $|{\bf k}|$
by $E_p = \sqrt{{\bf p}^2+m^2}$ in Eq.~(\ref{Integral}).
To evaluate the integration, we should expand the logarithm in the same way as
the four dimesional case. 
Individual integration is done by using the Bessel function as
\begin{align}
\frac{e^{\pm 2i \pi nq^{ij}}}{n} \int_0^\infty p^3 e^{-n L E_\mathrm{p}} dp
&= - \frac{e^{\pm 2i \pi nq^{ij}}}{n^2} \frac{\partial}{\partial L}
   \int_1^\infty m^3 (t^2-1) e^{-nLmt} dt.
\nonumber\\
&= - \frac{e^{\pm 2i \pi nq^{ij}}}{n^2} \frac{2^{3/2}}{\sqrt{\pi}}
   \frac{\partial}{\partial L}
   \Bigl[ \Bigl( \frac{m}{n L} \Bigr)^{3/2}
          K_{3/2} (n L m) \Bigr]
\nonumber\\
&= \frac{e^{\pm 2i \pi nq^{ij}}}{n} \frac{ 2^{3/2} m}{\sqrt{\pi}}
   \Bigl( \frac{m}{n L} \Bigr)^{3/2}
   K_{5/2}(n L m).
\end{align}
Then, the potential contributions from massive fermions are obtained by using the $K_{5/2}(x)$ as
\begin{align}
{\cal V}_f^\phi (N_{f}, m_{f})
&= \frac{ \sqrt{2} N_f ( m_{f}/L )^{5/2} }{\pi^{5/2}} 
   \sum_{i=1}^N \sum_{n=1}^\infty
   \frac{K_{5/2} (n m_{f} L)}{n^{5/2}} \cos[ 2 \pi n (q_i + \phi )],
\\
{\cal V}_a^\phi (N_{a}, m_{a})
&= \frac{ \sqrt{2} N_a ( m_{a}/L )^{5/2} }{\pi^{5/2}} 
   \sum_{i,j=1}^N \sum_{n=1}^\infty
   \Bigl( 1 - \frac{1}{N} \delta_{ij} \Bigr)
   \frac{K_{5/2} (n m_{a} L)}{n^{5/2}} \cos[ 2 \pi n (q_{ij} + \phi )].
\end{align}
This total one-loop effective potential in five dimensions again contains eight parameters including
the compactification scale $L$, the number of colors $N$, the fermion masses $m_{f}$, $m_{a}$, the number of flavors $N_{f}$, $N_{a}$, and the boundary conditions $\phi$ for two kinds of matter fields.
As with the four-dimensional case we keep $N=3$, then obtain phase diagram in $1/L$-$m_{a}$ space
with $m_{f}$, $N_{f}$, $N_{a}$ and $\phi$ fixed to several values.
The $1/L$-$m_{a}$ phase diagram obtained from this effective potential 
is valid at smaller $L$ or the weak coupling regime, 
and thus works to reveal properties of the exotic phases at weak coupling.

We here comment on treatment of parity pairs.
When we need a parity-even mass term in odd dimensions (e.g. $Z_2$ orbifolded theory), 
we need a set of parity pairs $\psi^{\pm}(-y)=\pm\psi(y)$ in the action \cite{MT1}.
Although our study does not take much care about such a case,
we note that the factor two appears in front of the fermion potentials
if the parity pair is introduced.
It is thus the same situation as doubled flavors within this setup.
In Sec.~\ref{sec:PS5}, we will see that the phase diagram changes depending on
the choice of the parity pair in five dimensions.

%%%%%%%%%%%%%PS4%%%%%%%%%%%%%%%%%%%

\section{Phase structure in four dimensions}
\label{sec:PS4}

In this section, we investigate the phase structure 
in the compacted-size $L^{-1}$ and quark-mass $m$ space 
for $SU(3)$ gauge theories on $R^{3}\times S^{1}$. 
Because of the relation $-q_{3}=q_{1}+q_{2}$, 
the vacuum structure is discussed just from the $(q_{1},q_{2})$ potential.
In the following we begin with investigation on vacuum and phase structures based on 
the one-loop potential~(\ref{eq_ep_pert}) for several choices of matters and boundary conditions.
By far, lattice QCD simulations in four dimensions with one compacted dimension have been done for aPBC fundamental, PBC fundamental, aPBC adjoint \cite{KL1} and PBC adjoint \cite{CD1} matters. 
There are no lattice study on systems including both fundamental and adjoint matters.
We also have no lattice corresponding results for five dimensional cases.
In the present study we mainly focus on the case with PBC adjoint, and with PBC adjoint and fundamental matters, and make comparison to the lattice result if exists. 
As shown in the above section, we make use of the other lattice results to fix model parameters
in PNJL models.

We note that we will sometimes consider one-flavor cases as
$(N_{f},N_{a})=(0,1)$ or $(N_{f},N_{a})=(1,1)$.
Since the anomaly is not implemented and difference of one- and multiple-flavors
just appears as an overall factor in the effective potential, 
our four-dimensional results are qualitatively unchanged even for
the two-flavor cases as $(N_{f},N_{a})=(0,2)$ or $(N_{f},N_{a})=(2,2)$.
Thus, these one-flavor examples are sufficient for our purpose of studying 
phase diagram qualitatively. (Things are different in five dimensions as shown in Sec.~\ref{sec:PS5}.)

In the appendix.~\ref{sec:VSF}, we summarize the potential minima for 
gauge-symmetry-preserved cases, which are not our main results.
We show the contour plots of the effective potentials for 
$(N_{f},N_{a})=(0,0)$ in Fig.~\ref{Fig_p_g_3D}, 
$(N_{f},N_{a})=(1,0)$ with the anti-periodic boundary condition
(aPBC) in Fig.~\ref{Fig_p_gfa_3D}, $(N_{f},N_{a})=(1,0)$ 
with the periodic boundary condition (PBC) in Fig.~\ref{Fig_p_gfp_3D}, 
and $(N_{f},N_{a})=(0,1)$ with aPBC in Fig.~\ref{Fig_p_gaa_3D}. 
We set $m_{f}=m_{a}=0$ in these cases.
In the pure-gauge and gauge+adjoint theories, the three degenerate vacua 
clearly reflects the $Z_{3}$ center symmetry while the introduction of fundamental fermions 
lift of the degeneracy, which means breaking of the center symmetry.
We note that the global minima are given by any of $(q_{1},q_{2})=(0,0),(1/3,1/3), (-1/3,-1/3)$,
and we have $q_{1}=q_{2}=q_{3}$ (mod 1) in the vacuum. 
It obviously shows that the $SU(3)$ symmetry remains in all these cases. 

 %%%%%%%%%%%%%%%% Fig %%
\begin{figure}[htbp]%[H]
\begin{center}
 \includegraphics[width=0.4\textwidth]{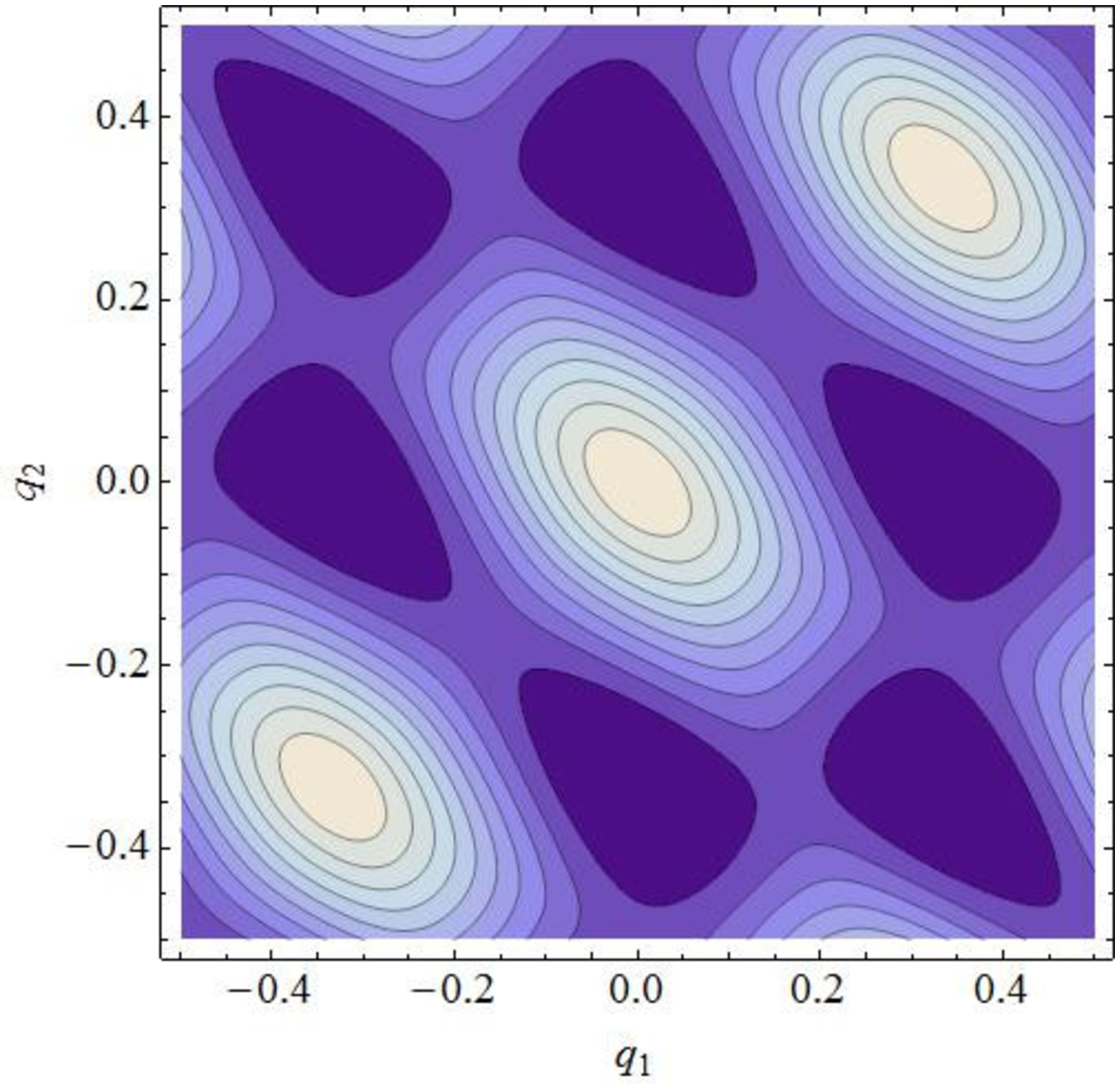}
 \includegraphics[width=0.4\textwidth]{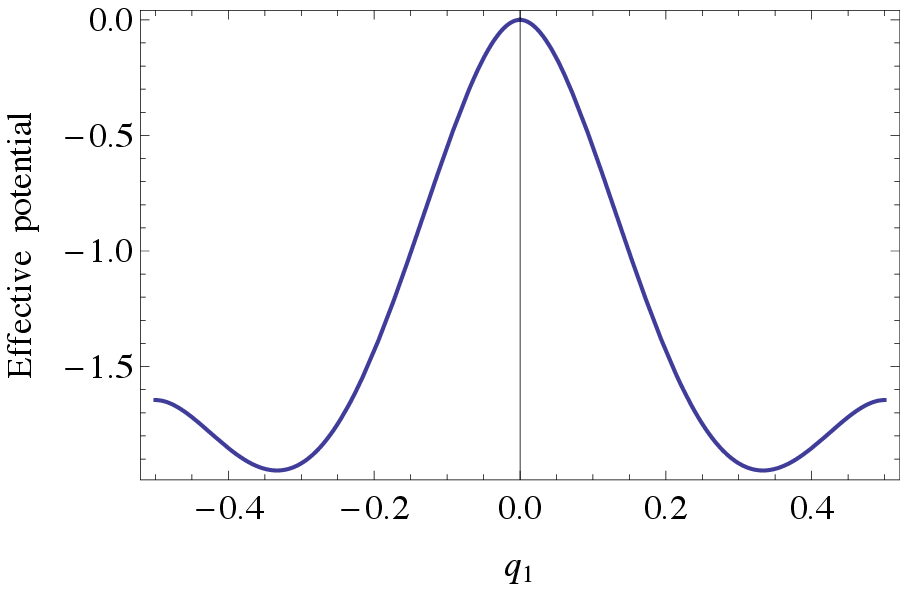}
\end{center}
\caption{
The one-loop effective potential of $SU(3)$ gauge theory on 
$R^{3}\times S^1$ with one adjoint fermion with PBC
$[  {\cal V}_g  + {\cal V}_a^{0}(N_{a}=1, m_a =0)] L^4$.
(Right) The contour plot as a function of $q_1$ and $q_2$.
Thicker region stands for deeper region of the potential.
(Left) The effective potential as a function of $q_1$ with $q_2=0$.
The global minima are located at $(q_{1},q_{2})=(\pm1/3,0)$.
}
\label{Fig_p_gap_3D}
\end{figure}
%%%%%%%%%%%%%%%%%%%%

Now, we shall look into the case with spontaneous gauge symmetry breaking.
We consider $(N_{f},N_{a})=(0,1)$ with PBC.
We note that this theory has exact center symmetry, and
all the phases, even the gauge-broken phase, 
should reflect this symmetry. 
Figure~\ref{Fig_p_gap_3D} shows the effective potential
$[ {\cal V}_g +{\cal V}_a^{0} ] L^4$.
The left contour plot is obviously different from the gauge-symmetric cases. 
Careful search shows that the minima are located at 
$(q_1,q_2)=(0,1/3)$, $(1/3,0)$, $(-1/3,1/3)$, $(-1/3,0)$, 
$(0,-1/3)$, $(1/3,-1/3)$.
It means the vacua are given by permutations of
$(q_1,q_2,q_3)=(0, 1/3, -1/3)$, and $SU(3)$ gauge symmetry 
is broken into $U(1)\times U(1)$.
This is the famous result, known as the Hosotani mechanism, where
the Aharonov-Bohm effect in the compacted dimension nontrivially
breaks gauge symmetry \cite{H1, H2, H3, H4}. 
We note that this situation is sometimes called ``re-confined phase" \cite{CD1}
since the color fundamental trace of the Polyakov loop 
$\Phi \equiv ({\rm Tr}_F~P) / N$ becomes zero.

\begin{figure}[htbp]%[H]
\begin{center}
 \includegraphics[width=0.23\textwidth]{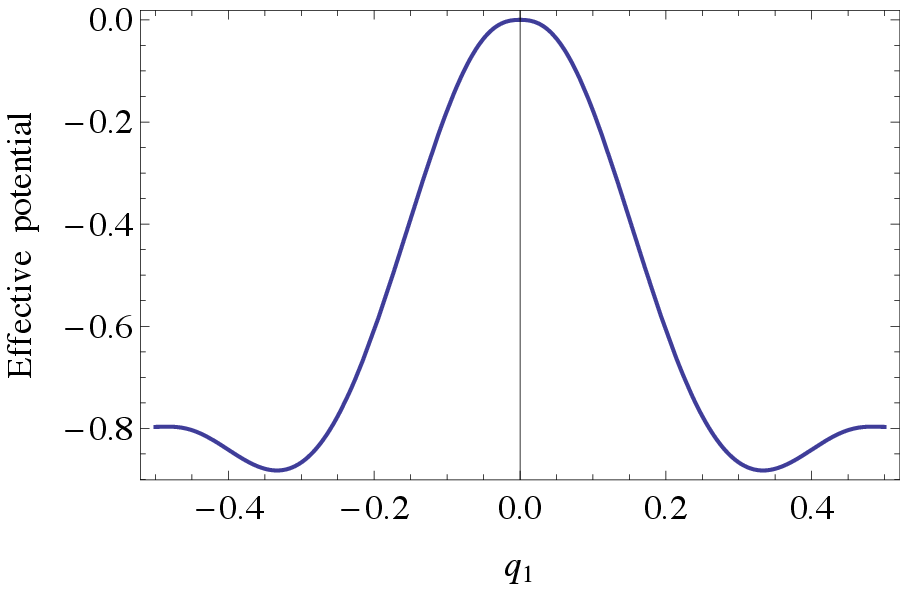}
 \includegraphics[width=0.23\textwidth]{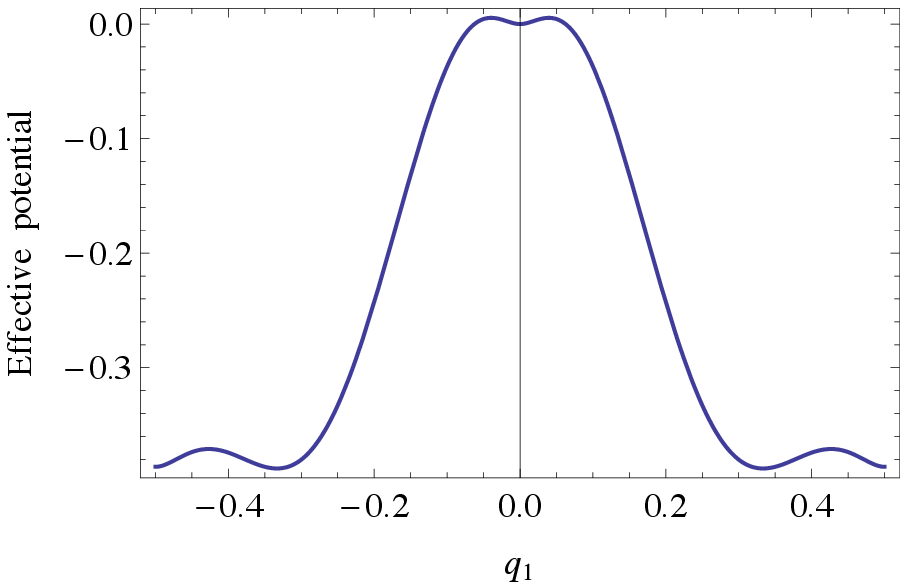}
 \includegraphics[width=0.23\textwidth]{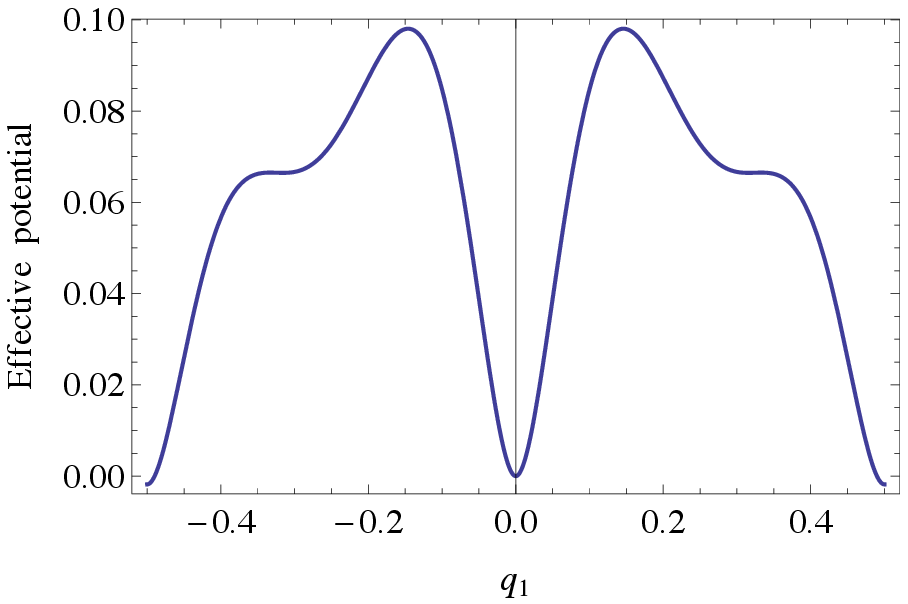}
 \includegraphics[width=0.23\textwidth]{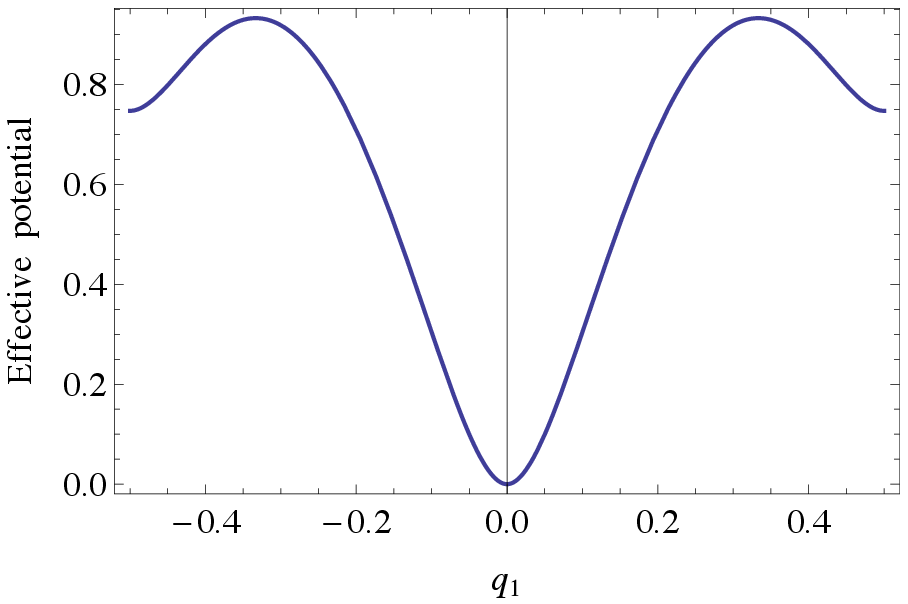}
\end{center}
\caption{ 
The one-loop effective potential of $SU(3)$ gauge theory on 
$R^{3}\times S^1$ with one PBC adjoint quark as a function of $q_1$ with $q_2=0$
$[ {\cal V}_g + {\cal V}_a^{0}] L^4$, for $m L=1.2$ (reconfined), 
$1.6$ (reconfined$\leftrightarrow$split), 
$2.0$ (split$\leftrightarrow$deconfined) and $3.0$ (deconfined).
}
\label{Fig_p_gapm_2D}
\end{figure}

\begin{figure}[htbp]%[H]
\begin{center}
 \includegraphics[width=0.4\textwidth]{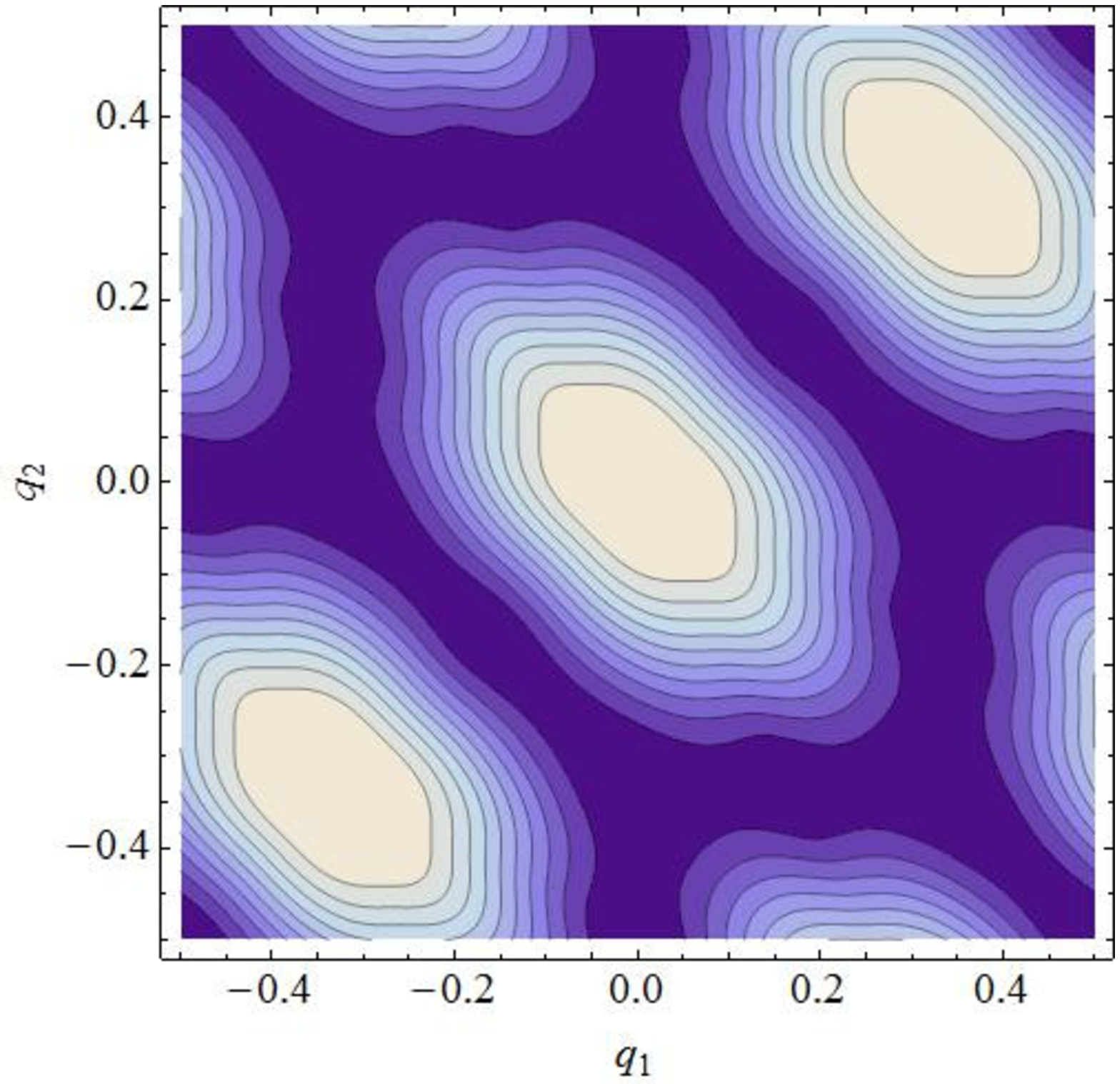}
 \includegraphics[width=0.4\textwidth]{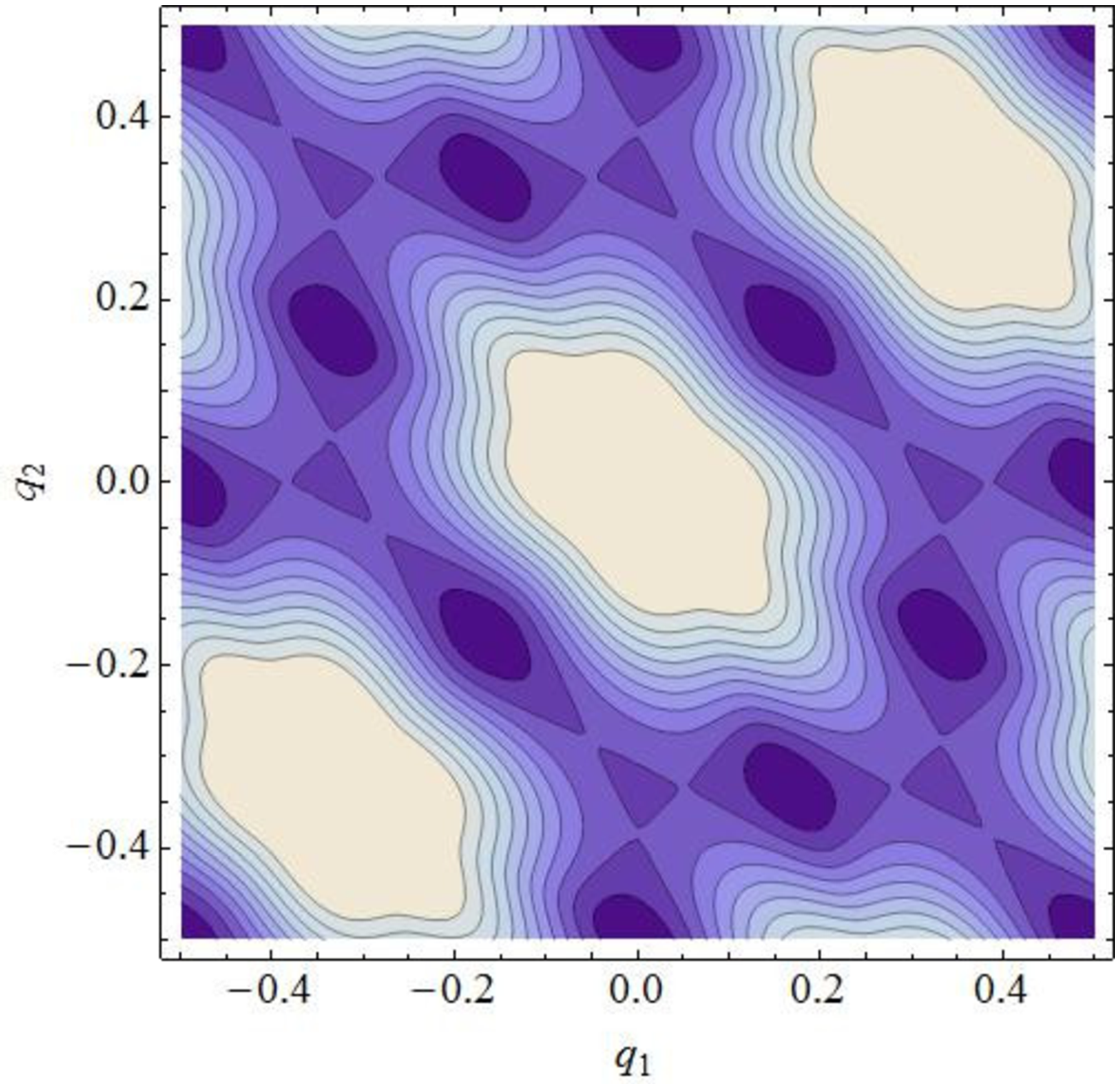}
\end{center}
\caption{
Contour plot of the one-loop effective potential of $SU(3)$ gauge theory on 
$R^{3}\times S^1$ with one PBC adjoint quark
$[ ( {\cal V}_g )_{pert} + {\cal V}_a^{0} ] L^4$,
for $m L=1.6$ and $1.8$ ($SU(2)\times U(1)$ split phase) 
as a function of $q_1$ and $q_2$.
Thicker region indicates deeper region of the potential.}
\label{Fig_p_gapm_3D}
\end{figure}

To study the phase diagram, we introduce nonzero quark mass.
Figure~\ref{Fig_p_gapm_2D} shows the effective potential
$[{\cal V}_g+{\cal V}_a^{0}(N_{a}, m_{a})]L^4$ as a function of $q_{1}$ with $q_{2}=0$
for $m L=1.2$, $1.6$, $2.0$ and $3.0$ from left to right panels ($m \equiv m_{a}$).
It is clearly seen that there is the first-order phase transition in
the vicinity of $m L=1.6$.
This is a transition between the re-confined phase and the other gauge-broken 
phase, which we call the ``split phase" \cite{CD1}.
The contour plots for $mL=1.6$ and $mL=1.8$ are shown in
Fig.~\ref{Fig_p_gapm_3D}.
The $mL=1.8$ case corresponds to the split phase.
The global minima in the split phase are given by
\begin{itemize}
\item $\mathrm{Im}~\Phi=0$ : 
      $(q_1,q_2)=(0,0.5)$, $(0.5,0.5)$, $(0.5,0)$,
\item $\mathrm{Im}~\Phi>0$ : $(q_1,q_2)=(1/6,-1/3)$, $(-1/3, 1/6)$, $(1/6,1/6)$,
\item $\mathrm{Im}~\Phi<0$ : $(q_1,q_2)=(1/3,-1/6)$, $(-1/6,-1/6)$, $(-1/6,1/3)$.
\end{itemize}
These results indicate that the vacuum in the first set is given by permutations of
$(q_1,q_2,q_3)=(0, 0.5, 0.5)$, and $SU(3)$ gauge symmetry 
is broken into $SU(2)\times U(1)$.
The vacua for other two sets are derived by the $Z_3$ transformation of
$(q_1,q_2,q_3)=(0, 0.5, 0.5)$.
From the viewpoint of phenomenological symmetry breaking, 
this phase and similar phases for large color numbers are the most preferable.

\begin{figure}[htbp]%[H]
\begin{center}
 \includegraphics[width=0.4\textwidth]{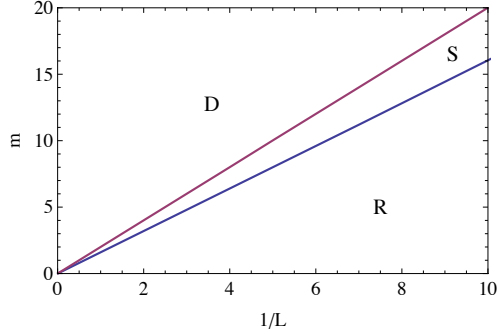}
\end{center}
\caption{$L^{-1}$-$m$ phase diagram for $SU(3)$ gauge theory on 
$R^{3}\times S^1$ with one PBC adjoint quark
based on one-loop effective potential. D stands for ``deconfined ($SU(3)$)", 
S for ``split ($SU(2)\times U(1)$)" and R for ``re-confined ($U(1)\times U(1)$)" phases.
Phase transitions are first-order.}
\label{4d_phase_p}
\end{figure}

\begin{figure}[htbp]%[H]
\begin{center}
 \includegraphics[width=0.4\textwidth]{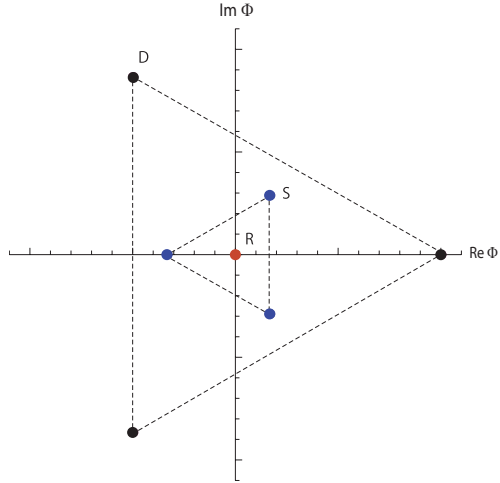}
\end{center}
\caption{
Schematic distribution plot of Polyakov loop $\Phi$ as a function of 
$\mathrm{Re}~\Phi$ and $\mathrm{Im}~\Phi$ for $SU(3)$ gauge theory on 
$R^{3}\times S^1$ with one PBC adjoint quark.
}
\label{Fig_pl_distribution}
\end{figure}

In Fig.~\ref{4d_phase_p} we depict the $L^{-1}$-$m$ phase diagram
with one PBC adjoint quark $(N_{f}, N_{a})=(0,1)$ 
based on the one-loop effective potential.
We note that, as $m$ appears as $m L$ in this one-loop effective potential,
we have scaling in the phase diagram and we can choose arbitrary mass-dimension unit.
Since we drop the non-perturbative effect in the gluon potential, 
we have no confined phase at small $L^{-1}$ (at low temperature).
The order of the three phases in Fig.~\ref{4d_phase_p} 
(deconfined $SU(3)$ $\to$ split $SU(2)\times U(1)$ $\to$ reconfined $U(1)\times U(1)$ 
from small to large $L^{-1}$.) 
is consistent with that of the lattice simulation except that we have no confined phase.
We note that all the critical lines in the figure are first-order.
In Fig.~\ref{Fig_pl_distribution} we depict a schematic distribution
plot of $\Phi$ in the complex plane.
It is obvious that each phase reflects $Z_{3}$ symmetry.
In the split phase, $\Phi$ takes nonzero values but in a different manner from the deconfined phase.
In the re-confined phase, we exactly have $\Phi=0$ with the vacuum breaking the gauge symmetry.
%%%%%%%%%%%%%%%% Fig %%
%%%%%%%%%%%%%%%%%%%%
%%%%%%%%%%%%%%%% Fig %%
%%%%%%%%%%%%%%%%%%%%
%%%%%%%%%%%%%%%% Fig %%
%%%%%%%%%%%%%%%%%%%%
%%%%%%%%%%%%%%%% Fig %%
%%%%%%%%%%%%%%%%%%%%

\begin{figure}[htbp]%[H]
\begin{center}
 \includegraphics[width=0.4\textwidth]{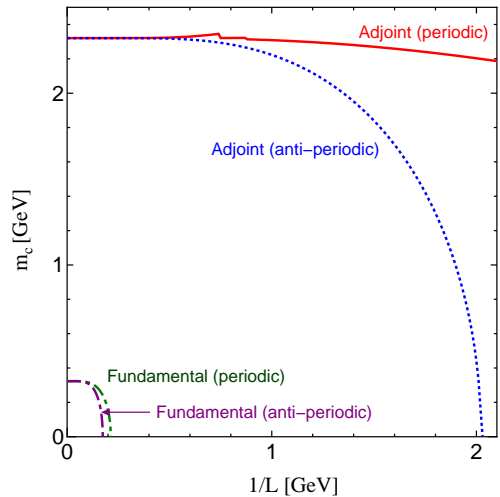}
\end{center}
\caption{Constituent mass $m_{c}$ as a function of $L^{-1}$
for $(N_{f},N_{a})=(2,0)$ with aPBC, $(N_{f},N_{a})=(2,0)$ with PBC,
$(N_{f},N_{a})=(0,2)$ with aPBC and $(N_{f},N_{a})=(2,0)$ with PBC
with the bare mass fixed as $m=0$ GeV.
We choose the parameter set as $\Lambda=0.63$ GeV and $g_{S}\Lambda^{2}=2.19$
for fundamental cases and $\Lambda=23.22$ GeV and $g_{S}\Lambda^{2}=0.63$
for adjoint cases.
}
\label{Fig_chiral}
\end{figure}

We next turn on the chiral sector and consider the PNJL effective potential (\ref{PNJL-like}).
We investigate chiral properties of $(N_{f},N_{a})=(0,2)$ $SU(3)$ gauge theory on $R^{3}\times S^{1}$.
\footnote{One may ask whether $N_{a}=2$ corresponds to the conformal window.
However, since we introduce nonzero quark mass in our study, the conformality, 
even if exists, is broken. }.
Before proceeding to the main topic, we discuss validity of the effective model (\ref{PNJL-like}) 
for the purpose of studying chiral properties at weak-coupling region.
Although we have no confined phase nor confined/deconfined phase transition in our model,
the chiral restoration associated with the phase transition is correctly reproduced in this model
for the known cases:  $(N_{f},N_{a})=(2,0)$ with aPBC, $(N_{f},N_{a})=(2,0)$ with PBC
and $(N_{f},N_{a})=(0,2)$ with aPBC. We depict behavior of the constituent mass
for these cases in Fig.~\ref{Fig_chiral}, where the chiral phase transition takes place at some point
for the three cases. The parameters are chosen so as to have the correct critical temperatures in finite-temperature $SU(3)$ gauge theory with aPBC quarks \cite{KL1, EHS1}, 
$\Lambda=0.63$ GeV and $g_{S}\Lambda^{2}=2.19$ for fundamental quarks 
and $\Lambda=23.22$ GeV and $g_{S}\Lambda^{2}=0.63$ for adjoint quarks \cite{NO1}.

\begin{figure}[htbp]%[H]
\begin{center}
 \includegraphics[width=0.4\textwidth]{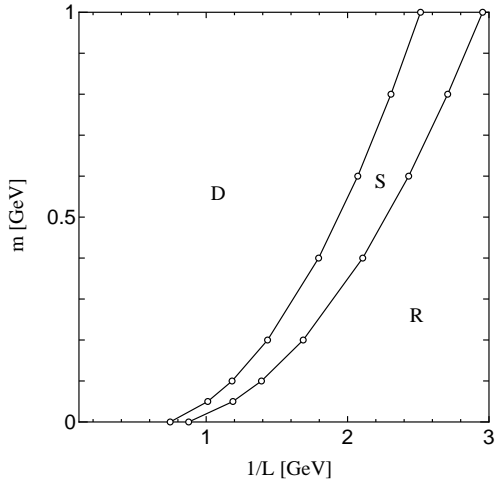}
\end{center}
\caption{$L^{-1}$-$m$ phase diagram for $SU(3)$ gauge theory on 
$R^{3}\times S^1$ with two-flavor PBC adjoint quarks $(N_{f},N_{a})=(0,2)$ 
based on the PNJL-type effective potential. D stands for ``deconfined", S for ``split"
and R for ``re-confined" phases. Critical lines are first-order.}
\label{Fig_chi_phase}
\end{figure}

We are now convinced that the model can work to study chiral properties, 
and we go on to the main topic, $(N_{f},N_{a})=(0,2)$ with PBC.
We calculate the PNJL effective potential in Eq.~(\ref{PNJL-like})
and search for the vacua for $q_{1}$, $q_{2}$ and $\sigma$. 
We depict the phase diagram for this case in Fig.~\ref{Fig_chi_phase}.
Due to nonzero constituent mass, the whole phase diagram is shifted,
and we have phase transitions even for a $m_{a}=0$ massless case.
We also note that the critical lines are curved due to the dimensionful parameters introduced.
Although our analysis is valid only for the weak-coupling regime,
the phase diagram is qualitatively consistent with that of the lattice simulations \cite{MO1,CD1}
except that ours have no confinement/deconfinement phase transition.
%%%%%%%%%%%%%%%% Fig %%
%%%%%%%%%%%%%%%%%%%%

\begin{figure}[htbp]%[H]
\begin{center}
 \includegraphics[width=0.4\textwidth]{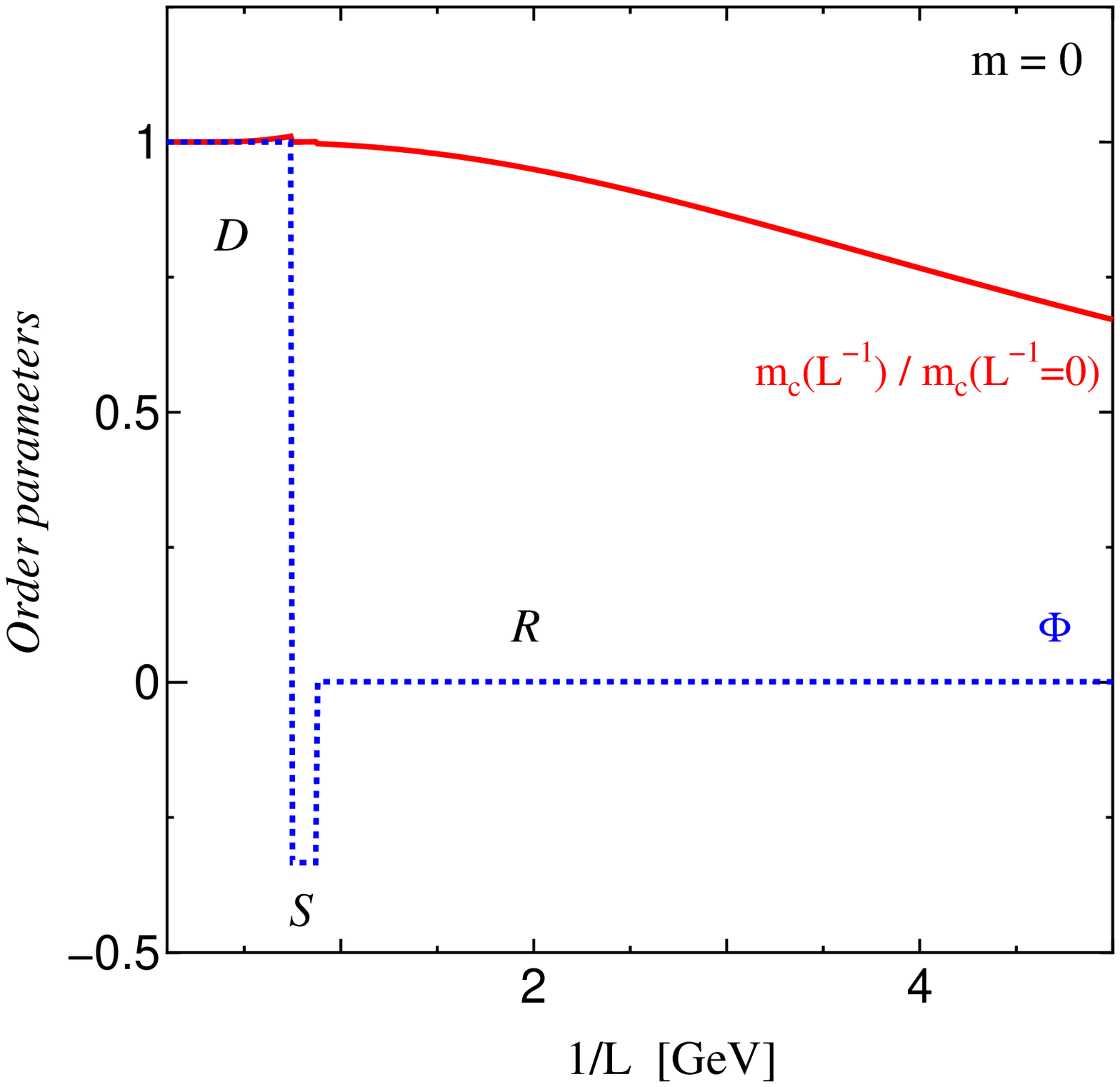}
 \includegraphics[width=0.4\textwidth]{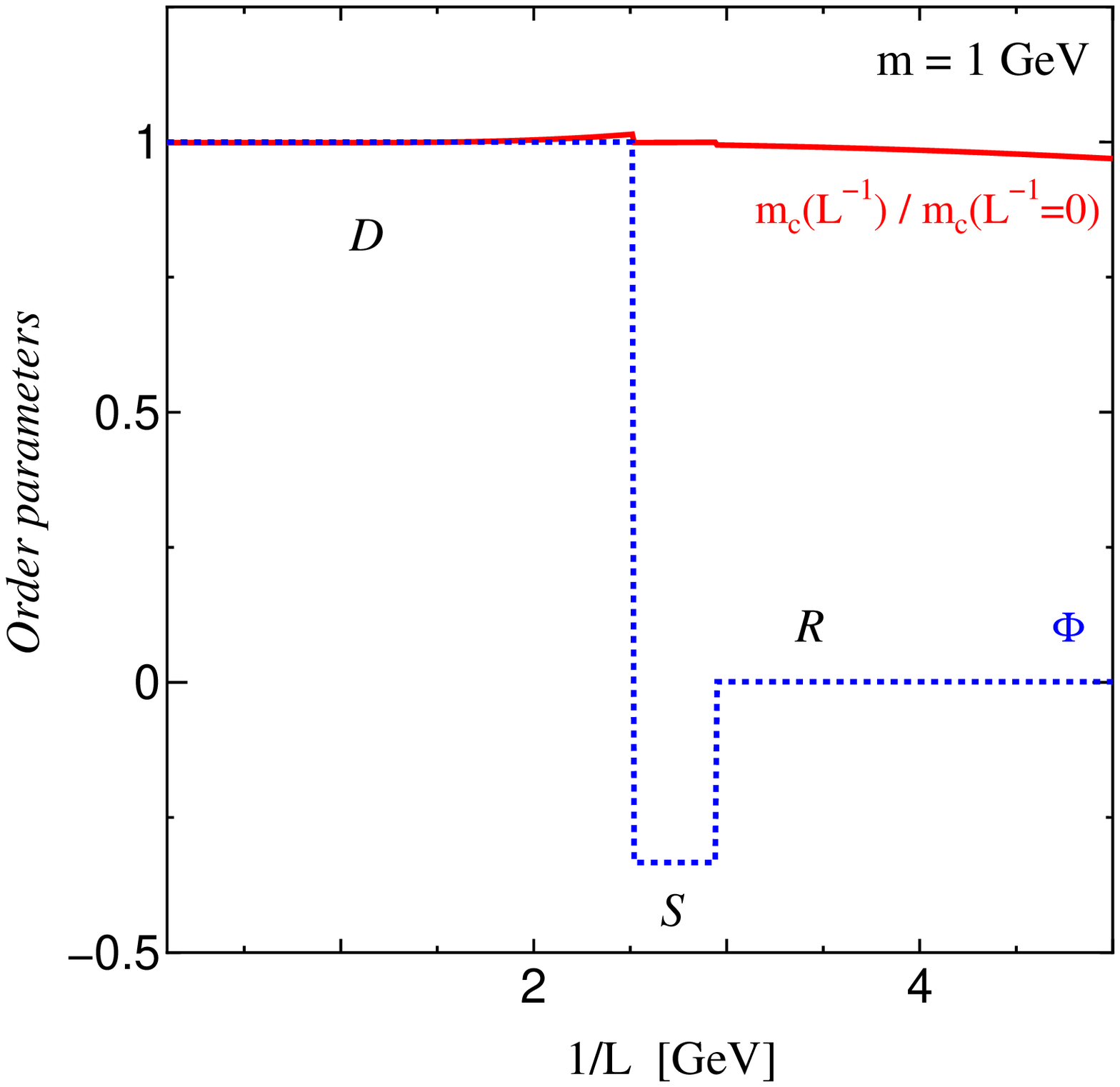}
\end{center}
\caption{
VEV of Polyakov loop $\Phi$ (blue dashed) with $q_2=0$ and
the constituent mass $m_{c}$ (red solid) as a function of $L^{-1}$ with the bare mass fixed as 
$m=0$ GeV (left) and $m=1$ GeV (right).
The constituent mass is normalized as $m_{c}(L^{-1})/m_{c}(L^{-1}=0)$.
The result indicates that the chiral symmetry is gradually restored without the clear phase transition. 
}
\label{Fig_chi_P}
\end{figure}

To look into chiral properties, we simultaneously depict the real part of VEV of Polyakov loop 
$\Phi$ and the constituent mass $m_{c}$ as a function of $L^{-1}$ with the bare mass fixed as 
$m=0$ GeV and $m=1$ GeV in Fig.~\ref{Fig_chi_P}.
We here normalize the constituent mass as $m_{c}(L^{1})/m_{c}(L^{-1}=0)$. 
It is notable that chiral condensate, or equivalently, constituent quark mass 
does not undergo a clear transition even for large values of $L^{-1}$, and
it gradually decreases.
We thus have no clear chiral restoration transition while chiral symmetry 
is gradually restored in this theory.
This result and the standard-PNJL result in \cite{NO1} are consistent with those
of the lattice simulation \cite{CD1}, which argues that the chiral restoration 
at weak coupling should occur at a quite small value of the compacted size $L$.
The other notable point is that the chiral condensate undergoes quite small transitions
coinciding with the deconfined/split and split/re-confined phase transitions.
(It can be seen better in Fig.~\ref{Fig_chiral} or the right panel for $m=1$ GeV in Fig.~\ref{Fig_chi_P}.)
This kind of the transition propagation is well studied in \cite{BCPG1, Kashiwa1}, but
they may be too small to be observed in the lattice simulations.
%%%%%%%%%%%%%%%% Fig %%
%%%%%%%%%%%%%%%%%%%%
%%%%%%%%%%%%%%%% Fig %%
%%%%%%%%%%%%%%%%%%%%

Let us briefly comment on non-perturbative deformation of the gluon potential 
discussed in Sec.~\ref{sec:EP}.
As a result, in Appendix.~\ref{sec:VSF}2 we show a result for the deformation (\ref{g_np}).
Indeed, with introducing the dimension parameter, for example $M=596$ MeV, 
the phase diagram in Fig.~\ref{4d_phase_p} is modified as Fig.~\ref{4d_phase_np}. 
Here the confined phase emerges at small $L^{-1}$ region, but 
it is connected with the re-confined phase through the small mass region.
This result clearly shows that the gauge symmetry is broken as $SU(3) \to U(1) \times U(1)$
even at zero-temperature or infinite-$L$. We note that the similar result with the same deformation in
the gluon potential is shown in Ref.~\cite{NO1}, where it is argued that the unified confined phase
implies the volume-independence of the confined phase structure.
For the other deformations, we have the same situation with explicit breaking of gauge symmetry.
For our purpose of clarifying and classifying phases of gauge symmetry, the deformations are not
appropriate although they may work as more phenomenological means.

\begin{figure}[htbp]%[H]
\begin{center}
 \includegraphics[width=0.4\textwidth]{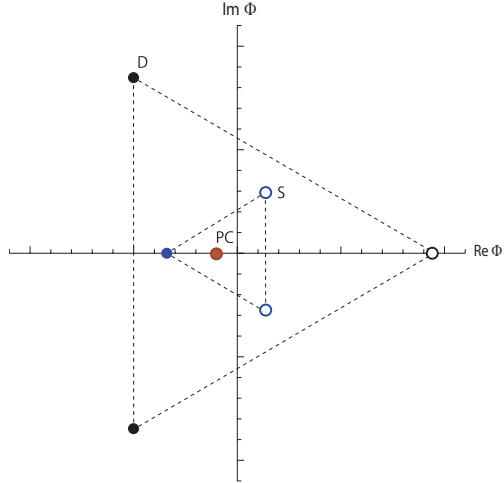}
\end{center}
\caption{
Schematic distribution plot of Polyakov loop $\Phi$ for $SU(3)$ gauge theory on 
$R^{3}\times S^1$ with one PBC adjoint and one PBC massless fundamental quarks 
$(N_{f},N_{a})=(1,1)$.
Points painted over stand for vacua in this case. $Z_{3}$ symmetry is broken, and some of the three minima are chosen as true vacua in deconfined and split phase. The Polyakov loop $\Phi$ in the 
pseudo-reconfined phase (PC) takes a nonzero and negative value.}
\label{Fig_pl_distribution2}
\end{figure}

\begin{figure}[htbp]%[H]
\begin{center}
 \includegraphics[width=0.4\textwidth]{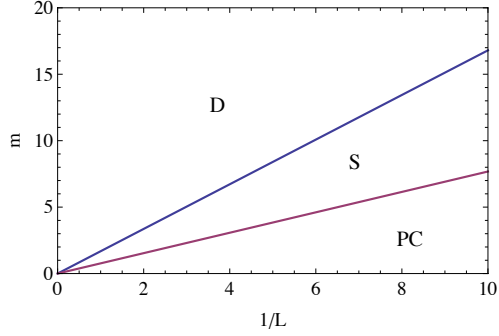}
\end{center}
\caption{
Phase diagram for $R^{3}\times S^1$ $SU(3)$ gauge theory 
with one PBC adjoint and one massless PBC fundamental quarks $(N_{f},N_{a})=(1,1)$ 
based on the one-loop effective potential. 
$m$ is the adjoint quark mass $m=m_{a}$ and $L^{-1}$ is the inverse of
the compacted size. PC stands for ``pseudo-reconfined" phase.}
\label{4d_phase_fp}
\end{figure}

We next consider $(N_{f},N_{a})=(1,1)$ with PBC.
We concentrate on the case with a massless fundamental quark, and 
the potential is given by 
${\cal V}_{g}+{\cal V}_{f}^{\phi}(N_{f}=1, m_f=0)+{\cal V}_a^{\phi}(N_{a}=1, m_a=m)$.
In this case, since the fundamental quark breaks the $Z_{3}$ center symmetry, 
the minima at ${\mathrm Re}~\Phi<0$ become true vacua in the deconfined and split phases
as shown in Fig.~\ref{Fig_pl_distribution2}. (Fundamental matter with PBC moves 
the vacua to ${\mathrm Re}~\Phi<0$ direction.)
In addition, we have no exact re-confiend phase since
$\Phi=0$ vacuum cannot be chosen because of the center symmetry breaking 
in Fig.~\ref{Fig_pl_distribution2}.
We term this unusual phase as ``pseudo-reconfined phase", where the $SU(3)$ 
is broken to $U(1)\times U(1)$ and $\Phi$ takes a nonzero and negative value.
The existence of this phase is consistent with the research on flavor-number dependence of 
gauge-symmetry-broken manners in Ref.~\cite{Hat1}.
In Fig.~\ref{4d_phase_fp}, we depict the phase diagram for the same case.
We emphasize that the split phase gets larger by introducing PBC fundamental quarks,
compared to Fig.~\ref{4d_phase_p}. 
It is because the center symmetry breaking chooses $(q_{1},q_{2},q_{3})=(0,0.5,0.5)$
and its permutations as true vacua among all the possible minima Fig.~\ref{Fig_pl_distribution2},
and makes it more stable than the center-symmetric case in Fig.~\ref{Fig_pl_distribution}.
%%%%%%%%%%%%%%%% Fig %%
%%%%%%%%%%%%%%%%%%%%
%%%%%%%%%%%%%%%% Fig %%
%%%%%%%%%%%%%%%%%%%%

\begin{figure}[htbp]%[H]
\begin{center}
 \includegraphics[width=0.4\textwidth]{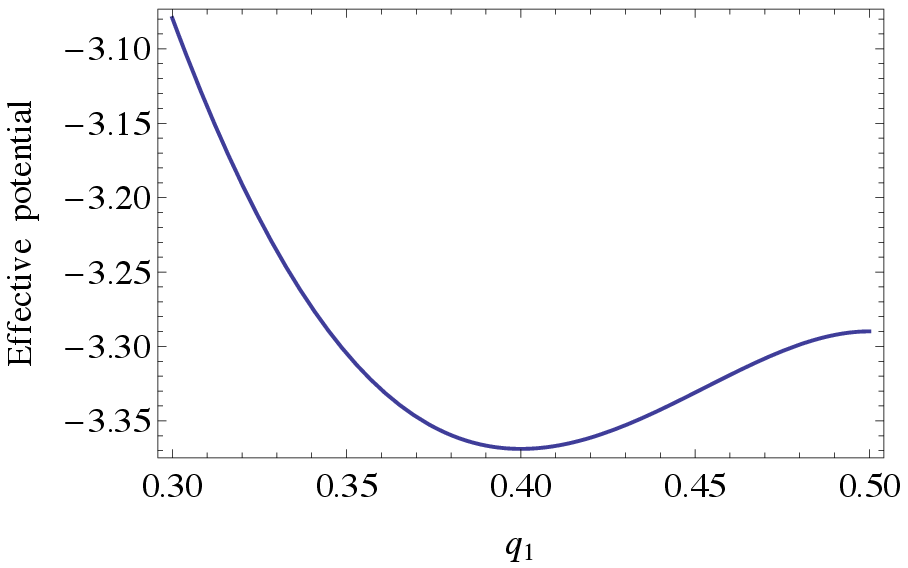}
 \includegraphics[width=0.4\textwidth]{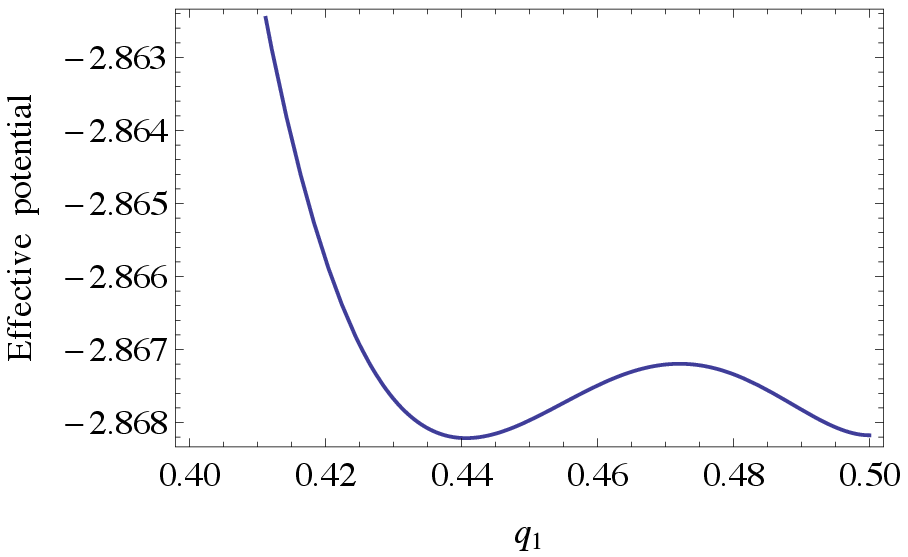}
\end{center}
\caption{Expanded effective potential of $R^{3}\times S^1$ $SU(3)$ gauge theory
with one PBC adjoint and one massless PBC fundamental quarks 
as a function of $q_1$ with $q_2=0$. 
Left one shows the case with the pseudo-reconfined phase ($m=m_{a}=0$), 
where we have the minimum at $(q_{1},q_{2})\sim(0.4,0)$.
Right one shows the first-order phase transition between the pseudo-reconfined and split phase
at $m_{a}L=0.77$.}
\label{skew-rec4D}
\end{figure}

To look into the details of pseudo-reconfined phase and the phase transition
to the split phase, we depict the expanded effective potential near the global minima 
as a function of $q_1$ with $q_2 =0$ in Fig.~\ref{skew-rec4D}.
The left panel shows the result for the massless adjoint quark ($m_{a}=m=0$),
which corresponds to the pseudo-confined phase.
The minimum is not located at $(q_1,q_2)=(0.5,0)$(split case) nor 
at $(q_1,q_2)=(1/3,0)$(re-confined case). 
In the pseudo-confined phase we totally have six minima for $q_1$ and $q_2$ as
$(q_{1},q_{2})\sim(0,0.4)$, $(0.4,0)$, $(-0.4,0.4)$, $(-0.4,0)$, 
$(0,-0.4)$, $(0.4,-0.4)$, which means that the vacua are given by
the permutation of $(q_{1},q_{2},q_{3})\sim(0,0.4,-0.4)$. 
The right panel shows the first-order phase transition between pseudo-confined and split phases.
Since the potential barrier at the phase transition is quite low, 
the fluctuation could break down clear phase transition.
For the cases with $(N_{f}, N_{a})=(1,2)$ and $(N_{f}, N_{a})=(1,3)$ where the flavor of adjoint quarks
is larger than that of fundamental quarks, the minima of the potential in the pseudo-reconfined phase
becomes deeper, and we can observe the first-order phase transition more distinctly.
%%%%%%%%%%%%%%%% Fig %%
%%%%%%%%%%%%%%%%%%%%

\begin{figure}[htbp]%[H]
\begin{center}
 \includegraphics[width=0.4\textwidth]{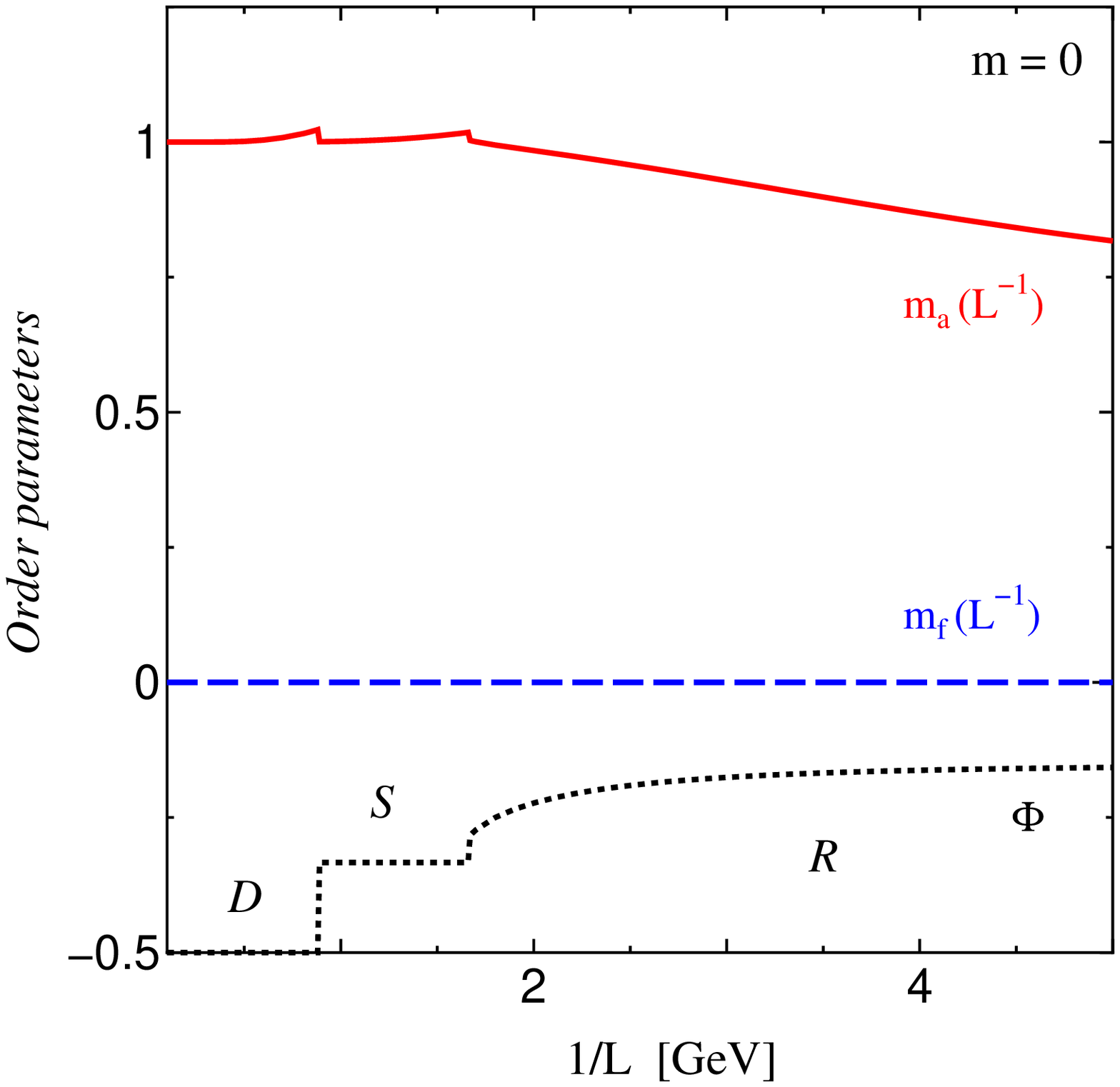}
 \includegraphics[width=0.4\textwidth]{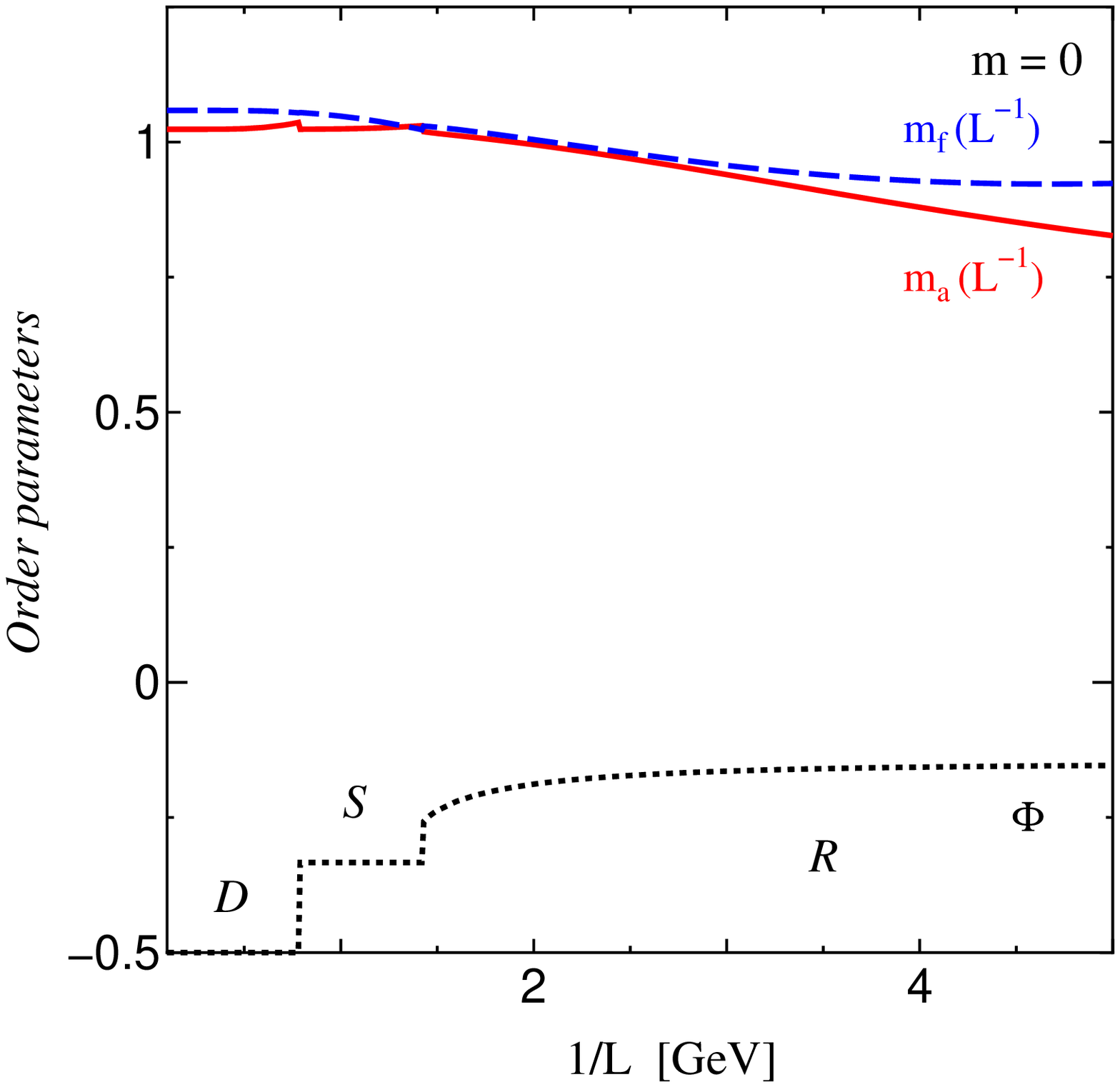}
\end{center}
\caption{
VEV of Polyakov loop $\Phi$ (black dotted),
the adjoint fermion constituent mass $m_{a}$ (red solid) and
the fundamental fermion constituent mass $m_{f}$ (blue dashed) 
as a function of $L^{-1}$ for the scenario A (left) and B (right).
The constituent masses are normalized as by $2.32$ GeV}
\label{Fig_chi_P1}
\end{figure}

The chiral sector can be introduced by extending to $(N_{f}, N_{a})=(2,2)$.
In this case we have two chiral sectors for fundamental and adjoint quarks, and
we have arbitrariness how to implement four-point interactions and choose relative parameters.
For example, we may consider the following forms of the four-point and eight-point interactions.
\begin{align}
 (g_{S})_{f}[(\bar{\psi}_{f}\psi_{f})^{2}&
            +(\bar{\psi}_{f}i\gamma_{5}{\vec \tau}\psi_{f})^{2}]
+(g_{S})_{a}[(\bar{\psi}_{a}\psi_{a})^{2}
            +(\bar{\psi}_{a}i\gamma_{5}{\vec \tau}\psi_{a})^{2}]
\nonumber\\
\,\,\,\,\,\,\,\,\,\,\,\,\,\,\,\,\,\,\,\,\,\,\,\,\,\,\,\,
&+(g_{S})_{fa}[\{(\bar{\psi}_{f}\psi_{f})^2
                +(\bar{\psi}_{f}i\gamma_{5}{\vec \tau}\psi_{f})^2 \}^2
               \{(\bar{\psi}_{a} \psi_{a})^2
                +(\bar{\psi}_{a}i\gamma_{5}{\vec \tau}\psi_{a})^2 \}^2],
\end{align}
where $\psi_{f}$ and $\psi_{a}$ stand for fundamental and adjoint quark fields, and
$(g_{S})_{f}, (g_{S})_{a}, (g_{S})_{fa}$ stand for fundamental, adjoint and mixing effective coupling.
Even if we fix a form of the four-point interactions, we still have no criterion on how 
to set the parameters since there is no lattice study on this case either for aPBC or PBC.
Thus we just show results of the chiral properties for two representative sets of the parameters.
In either set, $g_a$ and $\Lambda$ are set the same value used in
Fig. \ref{Fig_chiral} and the mixing term is set as $(g_{S})_{fa}= g_a/\Lambda^6$.
As the first case we consider $(g_{S})_f = R_{\it Fierz} \times (g_{S})_a$, 
where $R_{\it Fierz}$ is the coefficient
obtained from the Fierz transformation. We call it ''scenario A", where
the fundamental chiral condensate becomes zero for all the region of
$1/L$ while the adjoint chiral condensate has the qualitatively similar
behavior to $(N_{f}, N_{a})=(0,2)$ case as shown in
left-panel of Fig.~\ref{Fig_chi_P1}.
As the second case we consider $(g_{S})_f = (8/3) \times (g_{S})_a$, which we call scenario B.
In this case the fundamental chiral condensate has nonzero value at $1/L=0$
as shown in the right-panel of Fig.~\ref{Fig_chi_P1}.
In this case, unlike the adjoint chiral condensate, the fundamental chiral condensate never reacts to
the gauge symmetry phase transitions and the decreasing behavior becomes relatively gentle
at large $1/L$ ($1/L \sim 5$ GeV).
We consider that either of the scenarios of chiral properties for $(N_{f}, N_{a})=(2,2)$
will be detected in the on-going lattice simulation.

%%%%%%%%%%%%%%%% Fig %%
%%%%%%%%%%%%%%%%%%%%
%For $(g_{S})_{f}=, (g_{S})_{a}=, (g_{S})_{fa}=, \Lambda=$, which we call scenario A, 
%the fundamental chiral condensate becomes zero for all the regions of $1/L$ while
%the adjoint chiral condensate has the qualitatively similar behavior to $(N_{f}, N_{a})=(0,2)$ case
%as shown in Fig.~\ref{Fig_chi_P1}.
%For $(g_{S})_{f}=, (g_{S})_{a}=, (g_{S})_{fa}=, \Lambda=$, which we call scenario B,
%the fundamental chiral condensate has nonzero value at $1/L=0$, and
%gradually decreases with $1/L$ increasing in Fig.~\ref{Fig_chi_P2}.

Besides the cases we have shown above, we also have other interesting choices of matters.
As shown in \cite{Kouno1}, the $SU(3)$ gauge theory with 
three fundamental flavors with flavored twisted boundary
conditions $\phi=0, \,1/3,\, 2/3$ can accidentally keep the $Z_{3}$ center symmetry.
For example, in the case of $(N_{f},N_{a})=(3,1)$ with the flavored twisted boundary conditions 
$\phi=0,\, 1/3,\, 2/3$ for fundamental fermions and PBC for the adjoint fermion, 
the distribution plot of $\Phi$ becomes symmetric as Fig.~\ref{Fig_pl_distribution}
and we have exact re-confined phase although the theory contains fundamental quarks. 
Moreover, in \cite{Kouno2}, the present authors and collaborators 
discuss the $SU(3)$ gauge theory with $(N_{f},N_{a})=(3,0)$ 
with the flavored twisted boundary conditions $\phi=0,\, 1/3,\, 2/3$
can lead to the spontaneous gauge symmetry breaking without adjoint quarks.
This case is also fascinating as a future research topic.

%%%%%%%%%%%%PS5%%%%%%%%%%%%%%%%%%%

\section{Phase structure in five dimensions}
\label{sec:PS5}

In this section, we discuss the vacuum and phase structure in $SU(3)$ 
gauge theories on $R^{4}\times S^{1}$. Since the five-dimensional gauge theory 
is not renormalizable, we cannot discuss its non-perturbative aspects 
in a parallel way to the four-dimensional case. 
Thus we will concentrate on the one-loop part of the effective theory, 
and will not consider the contribution from the chiral sector. 
It is sufficient for our purpose of investigating the phase structure
at weak-coupling.

\begin{figure}[htbp]%[H]
\begin{center}
 \includegraphics[width=0.3\textwidth]{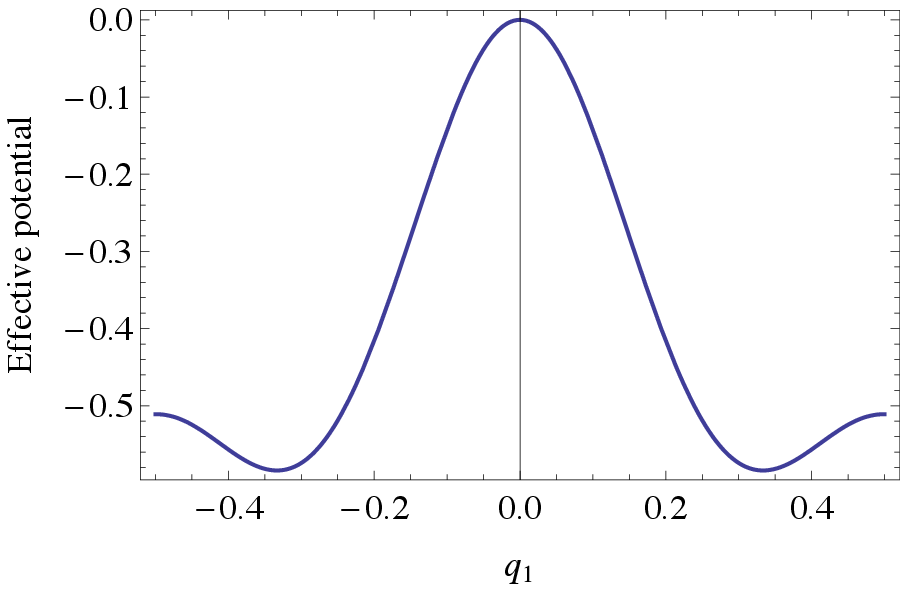}
 \includegraphics[width=0.3\textwidth]{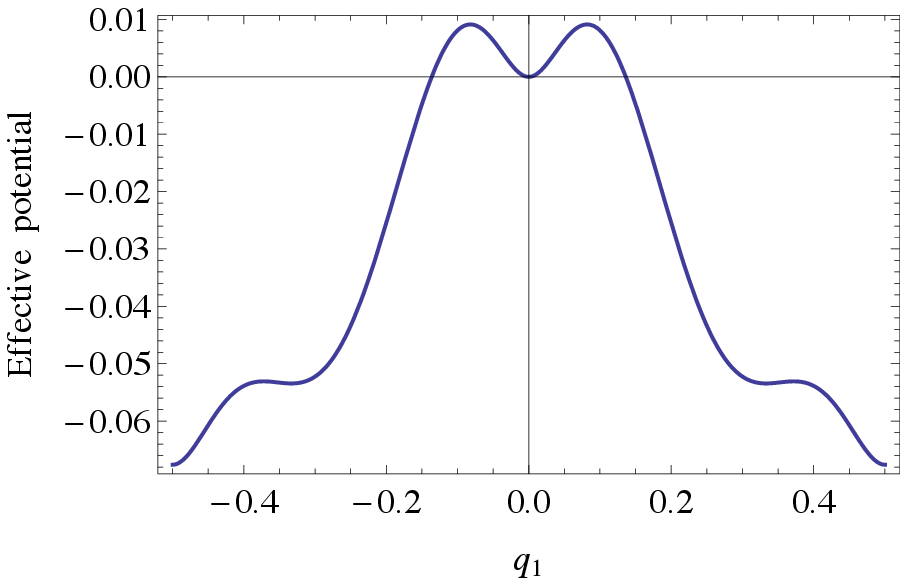}
 \includegraphics[width=0.3\textwidth]{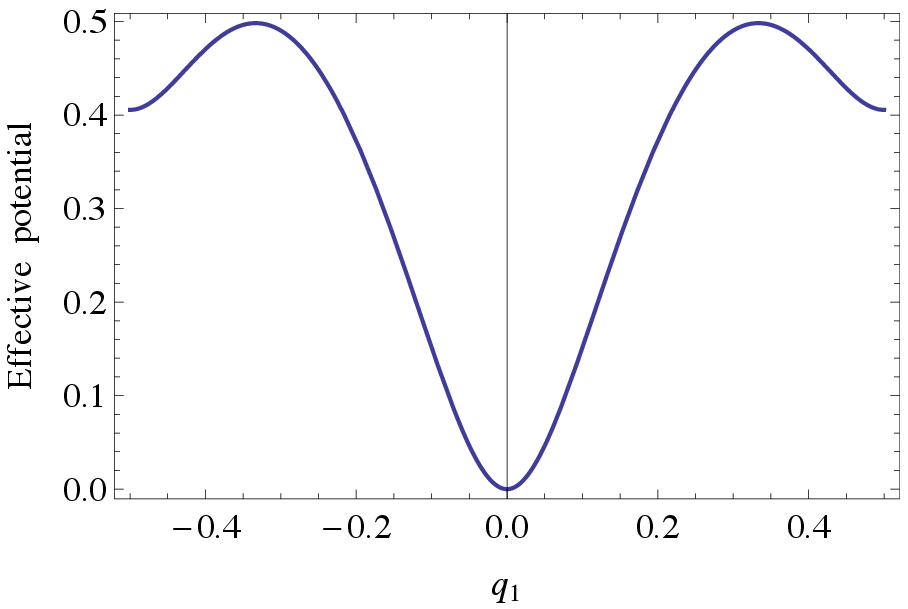}
\end{center}
\caption{The one-loop effective potential of $SU(3)$ gauge theory on 
$R^{4}\times S^1$ with one PBC adjoint quark as a function of $q_1$ with $q_2=0$
$({\cal V}_g+{\cal V}_a^{0}(1, m)) L^5$.
Depicted for $m L=0.5$ (re-confined), 
$1.3$ (split) and $2.0$ (deconfined).
}
\label{Fig_p_gapm_D5_2D}
\end{figure}

\begin{figure}[htbp]%[H]
\begin{center}
 \includegraphics[width=0.4\textwidth]{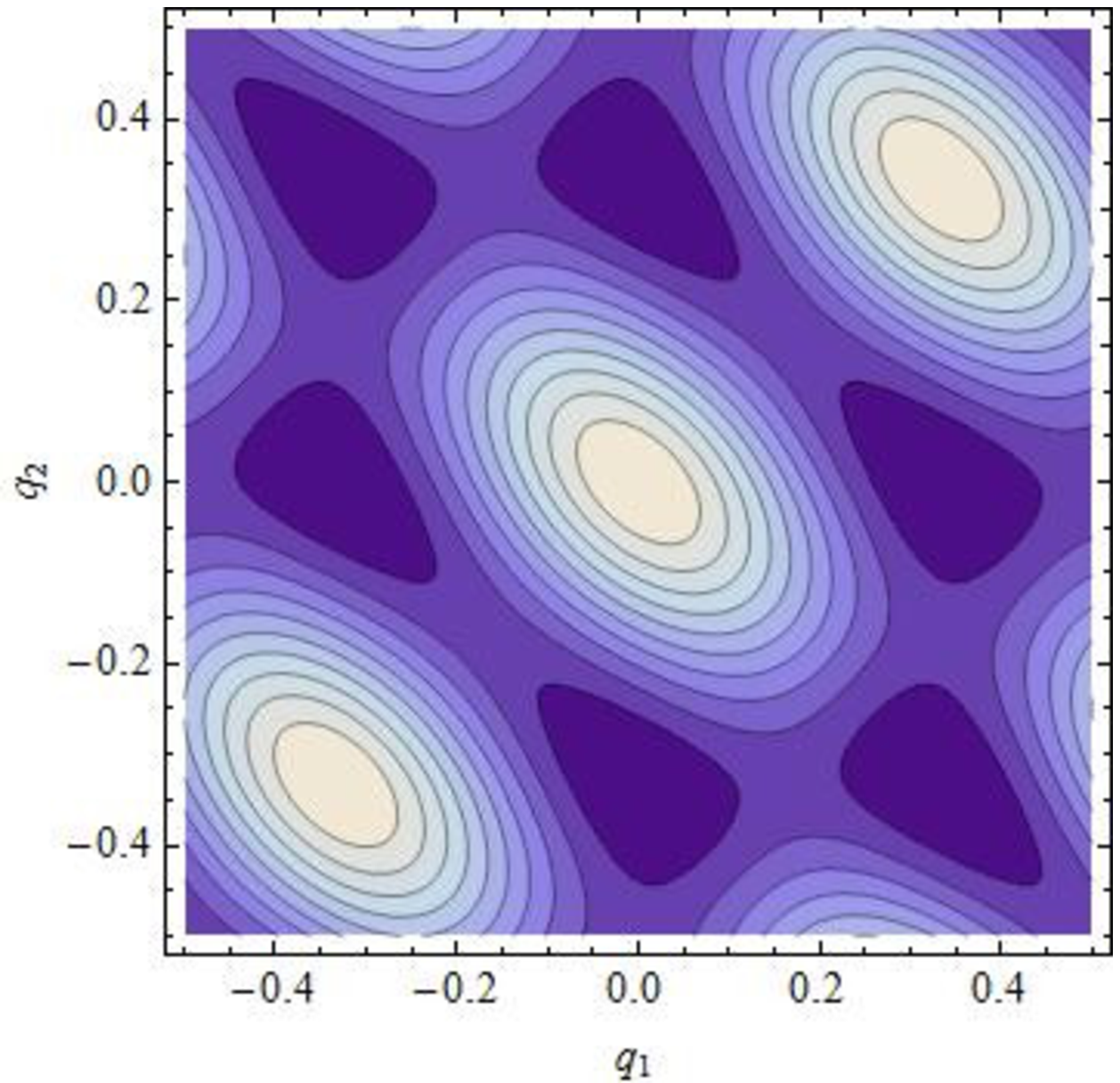}
 \includegraphics[width=0.4\textwidth]{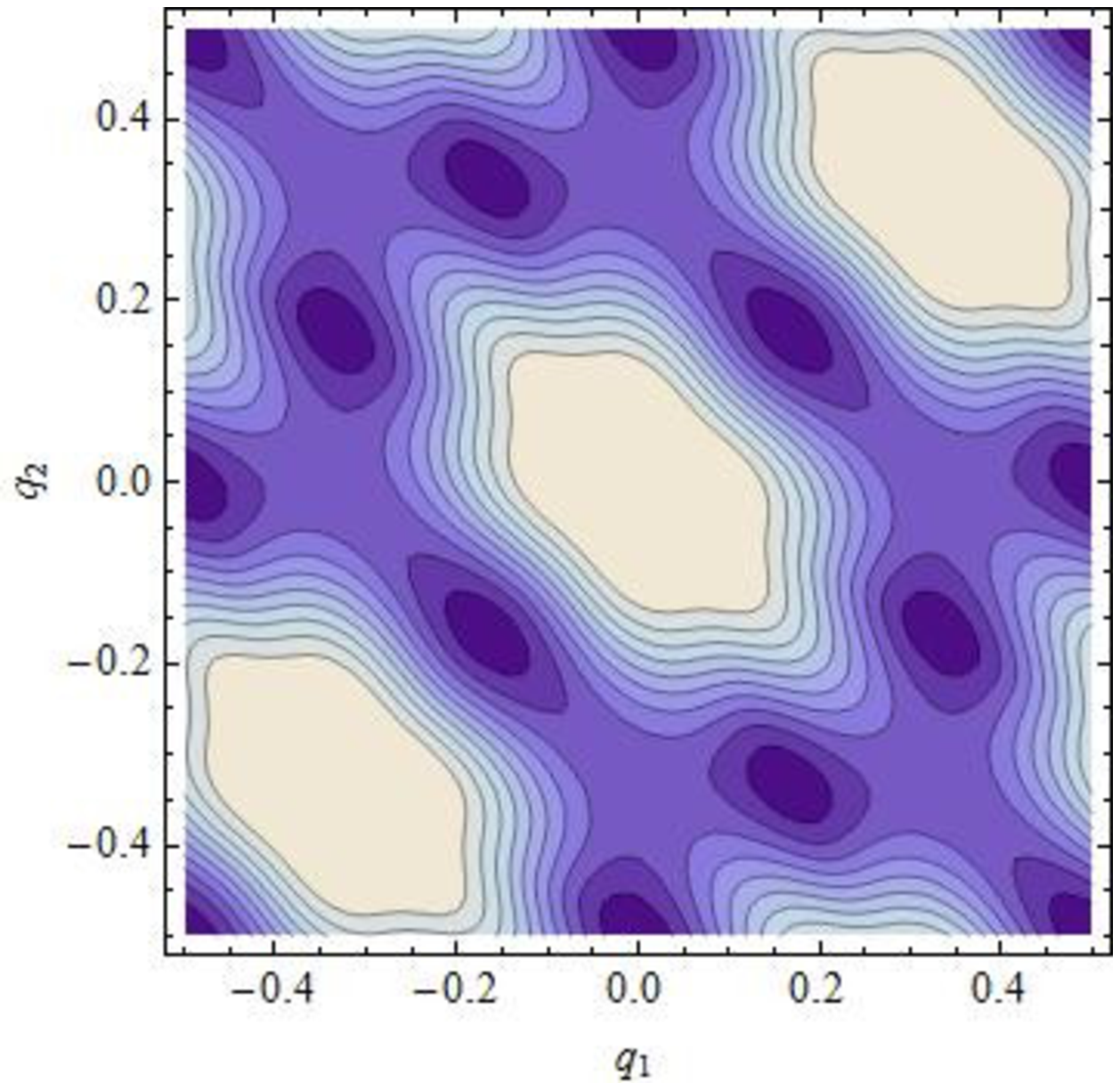}
\end{center}
\caption{
Contour plot of the one-loop effective potential of $SU(3)$ gauge theory on 
$R^{4}\times S^1$ with one PBC adjoint quark 
$({\cal V}_g+{\cal V}_a^{0}(N_{a}=1,m_{a}=m)) L^5$ 
for $mL=0.5$ and $1.3$ as a function of $q_1$ and $q_2$.
}
\label{Fig_p_gfapm_D5_3D}
\end{figure}

\begin{figure}[htbp]%[H]
\begin{center}
 \includegraphics[width=0.4\textwidth]{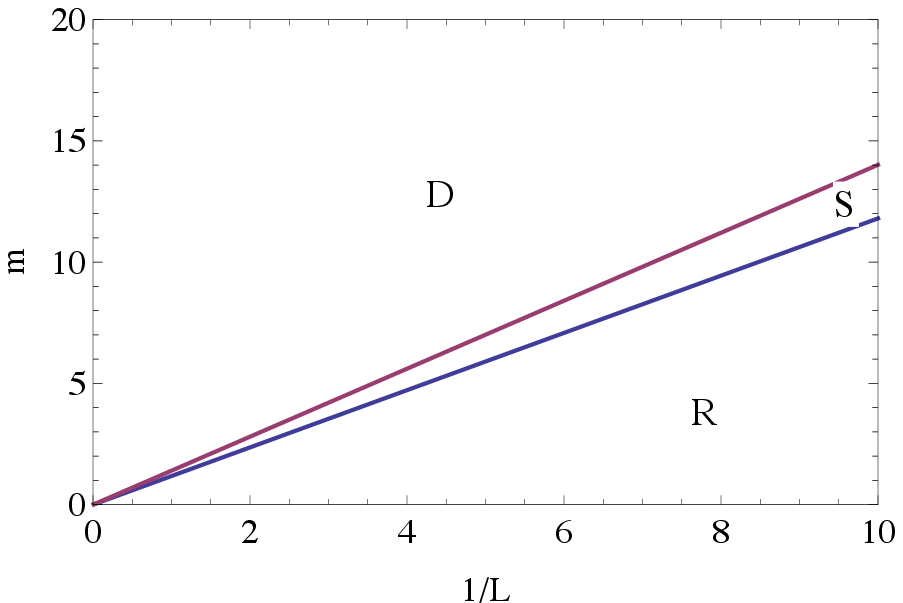}
\end{center}
\caption{
$L^{-1}$-$m_{a}$ phase diagram for $SU(3)$ gauge theory on 
$R^{4}\times S^1$ with one PBC adjoint quark  based on
the one-loop effective potential. 
}
\label{Fig_PD_gap}
\end{figure}

We consider the five-dimensional one-loop effective potential 
$[  {\cal V}_g + {\cal V}_{f}^{0}(N_{f},m_{f}) + {\cal V}_a^{0}(N_{a},m_{a}) ] L^5$.
We first look at the case with $(N_{f},N_{a})=(0,1)$ with PBC.
The qualitative properties are common with the four-dimensional case.
Figure~\ref{Fig_p_gapm_D5_2D} shows the effective potential 
 $[{\cal V}_g+{\cal V}_a^{0}(N_{a}=1, m)]L^5$ as a function of $q_{1}$ with $q_2=0$ fixed.
The adjoint quark mass is set as $m L=0.5$, $1.3$ and $2.0$ from left to right panels. 
The three cases correspond to re-confined (left), split (center), and deconfined (right) phase.
The contour plots at $m L=0.5$(reconfined) and $mL =1.3$(split) are depicted 
in Fig.~\ref{Fig_p_gfapm_D5_3D}.
The manners of $SU(3)$ symmetry breaking and
the distribution of Polyakov line $\Phi$ in each phase are the same as the four-dimensional case;
$SU(3)\to U(1)\times U(1)$ in the re-confined phase and $SU(3)\to SU(2)\times U(1)$ in the split phase.
We depict the phase diagram in $L^{-1}$-$m$ for this case in Fig.~\ref{Fig_PD_gap}.
The qualitative configuration of the three phases in the phase diagram
is indifferent from the four-dimensional case although the split phase
is smaller in five dimensions.
%%%%%%%%%%%%%%%% Fig %%
%%%%%%%%%%%%%%%%%%%%
%%%%%%%%%%%%%%%% Fig %%
%%%%%%%%%%%%%%%%%%%
%%%%%%%%%%%%%%%%%%%%

\begin{figure}[htbp]%[H]
\begin{center}
 \includegraphics[width=0.4\textwidth]{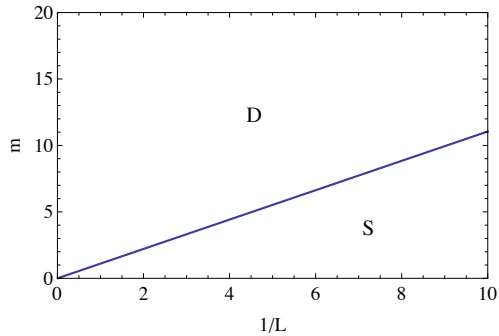}
\end{center}
\caption{
$L^{-1}$-$m_{a}$ phase diagram for $SU(3)$ gauge theory on $R^{4}\times S^{1}$ 
with one adjoint and one massless fundamental quarks ($(N_{f},N_{a})=(1,1)$, $m_{f}=0$, PBC) 
based on the one-loop effective potential. 
}
\label{Fig_PD_gapfp}
\end{figure}

\begin{figure}[htbp]%[H]
\begin{center}
 \includegraphics[width=0.4\textwidth]{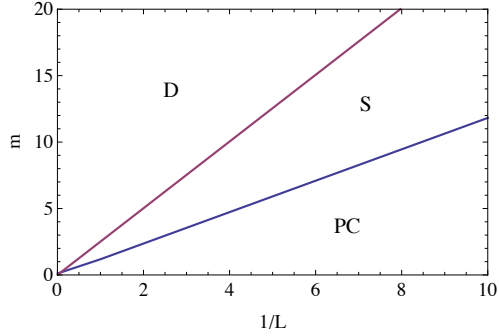}
\end{center}
\caption{
$L^{-1}$-$m_{a}$ phase diagram for $SU(3)$ gauge theory on $R^{4}\times S^{1}$ 
with a set of parity pairs of $(N_{f},N_{a})=(1,1)$, or equivalently $(N_{f},N_{a})=(2,2)$,
based on the one-loop effective potential.
}
\label{Fig_PD_gapfp2}
\end{figure}

\begin{figure}[htbp]%[H]
\begin{center}
 \includegraphics[width=0.4\textwidth]{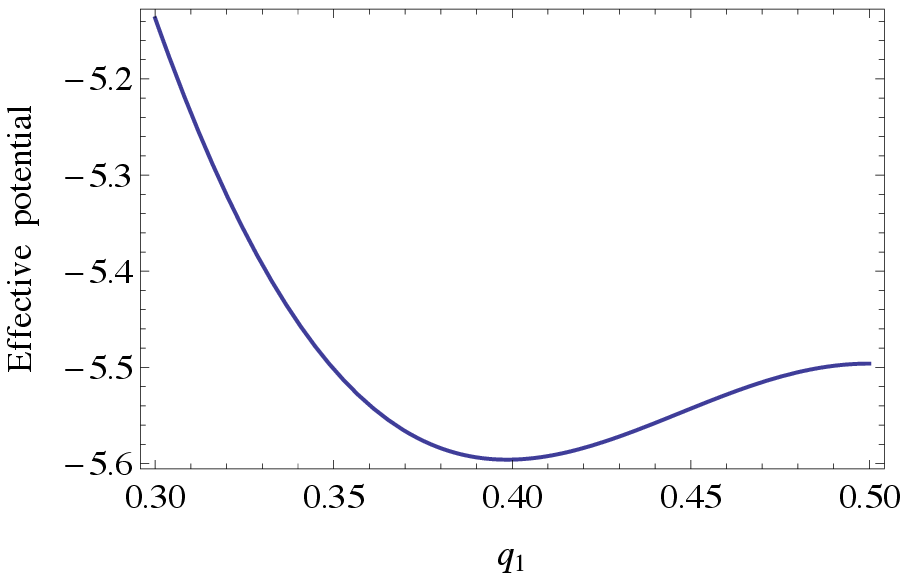}
 \includegraphics[width=0.45\textwidth]{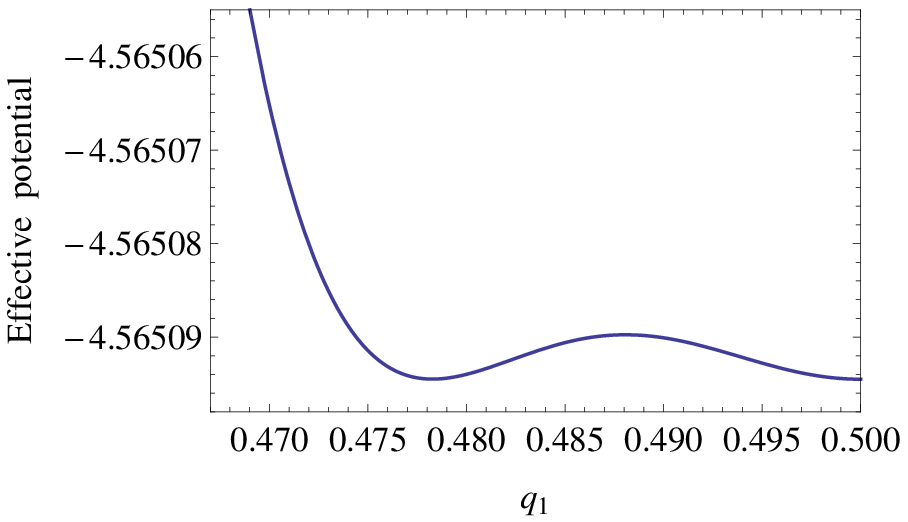}
\end{center}
\caption{Expanded 5D effective potential of $SU(3)$ gauge theory on $R^{4}\times S^{1}$ 
with a set of parity pairs of $(N_{f},N_{a})=(1,1)$ or $(N_{f},N_{a})=(2,2)$ 
as a function of $q_1$ with $q_2=0$. 
Left one shows the pseudo-confined phase ($m=m_{a}=0$), 
where we have the minimum at $(q_{1},q_{2})\sim(0.4,0)$.
Right one shows the first-order phase transition between 
the pseudo-reconfined and split phase at $m_{a}L=1.18$.}
\label{skew-rec5D}
\end{figure}

We next consider the case for $(N_{f},N_{a})=(1,1)$ with PBC.
We again concentrate on $m_{f}=0$.
Due to explicit breaking of $Z_{3}$ center symmetry, 
$\Phi<0$ minima are chosen as vacua in the deconfined and split phases 
in the same way as the four dimensional case (Fig.~\ref{Fig_pl_distribution2}).
It is notable that the split phase is again enlarged by introducing fundamental fermions, 
but more effectively in five dimensions.
Enhancement of the split phase due to the fundamental matter in five dimensions 
is sensitive to choice of parity pairs, or equivalently, choice of the number of flavors.
For $(N_{f},N_{a})=(1,1)$ without the parity pair, 
the pseudo-confined phase disappears and the split phase becomes the unique
gauge-broken phase as shown in Fig.~\ref{Fig_PD_gapfp}.
This result is consistent with that of the massless case ($m_{a}=0$) in Ref.~\cite{H3, H4}.
On the other hand, when we introduce parity pairs for $(N_{f}, N_{a})=(1,1)$ or equivalently
consider $(N_{f},N_{a})=(2,2)$, 
the split phase becomes wider than that in Fig.~\ref{Fig_PD_gap} 
but the pseudo-reconfined phase still survives as shown in Fig.~\ref{Fig_PD_gapfp2}. 
In the five-dimensional pseudo-confined phase we again have six minima
 for $q_1$ and $q_2$ as $(q_{1},q_{2})\sim(0,0.4)$, $(0.4,0)$, $(-0.4,0.4)$, $(-0.4,0)$, 
$(0,-0.4)$, $(0.4,-0.4)$, which indicates that the minima are given by
the permutation of $(q_{1},q_{2},q_{3})\sim(0,0.4,-0.4)$. 
We depict the expanded effective potential in Fig.~\ref{skew-rec5D}.
The left panel shows the massless case ($m_{a}L=0$), 
which corresponds to the pseudo-reconfined phase.
The right panel shows the first-order phase transition 
between the pseudo-reconfined and split phases ($m_{a}L=1.18$).
In the cases with $(N_{f},N_{a})=(1,2)$ or $(N_{f},N_{a})=(1,3)$, the potential minima 
in the pseudo-reconfined phase becomes deeper and the phase transition gets more distinct.
%%%%%%%%%%%%%%%% Fig %%
%%%%%%%%%%%%%%%%%%%%
%%%%%%%%%%%%%%%% Fig %%
%%%%%%%%%%%%%%%%%%%%
 %%%%%%%%%%%%%%%% Fig %%
%%%%%%%%%%%%%%%%%%%%

%The difference of $(N_{f},N_{a})=(1,1)$ and $(2,2)$ can 
%be understood by using Fig.~\ref{Fig_pl_distribution2}.
%Let us fix $N_{a}$ and increase $N_{f}$ from zero.
%Then, $\Phi$ for the pseudo-reconfined vacuum moves to ${\mathrm Re} \Phi<0$ direction
%while $\Phi$ for the split vacuum remains at the same point in ${\mathrm Re} \Phi<0.
%At some number of $(N_{f})_{c}$, the pseudo-reconfined $\Phi$ reaches the
%split $\Phi$. This is the point that the split phase overcomes the pseudo-reconfined phase.
%This critical $(N_{f})_{c}$ depends on $N_{a}$ and dimensions.
%What we found above is that $(N_{f})_{c}>1$ for $N_{a}=1$ in five dimensions
%while $(N_{f})_{c}<2$ for $N_{a}=2$ in five dimensions. 

\begin{figure}[htbp]%[H]
\begin{center}
 \includegraphics[width=0.4\textwidth]{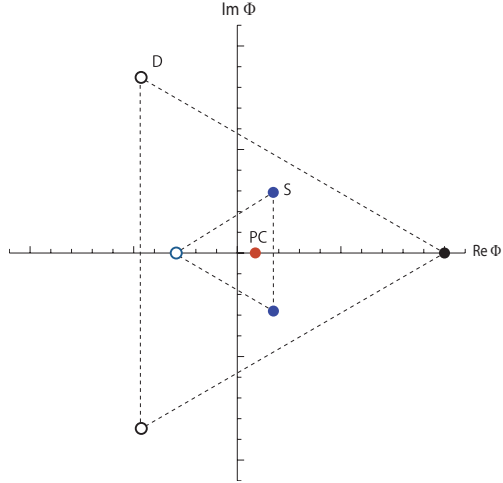}
\end{center}
\caption{
Schematic distribution plot of Polyakov loop $\Phi$ for SU(3) gauge theory
with one PBC adjoint and one aPBC fundamental quarks.
Points painted over stand for vacua in this case. 
$Z_{3}$ symmetry is broken in the different manner from Fig.~\ref{Fig_pl_distribution2}.}
\label{Fig_pl_distribution3}
\end{figure}

In the end of this section we comment on the other aspect of gauge theory
with a compacted dimension.
If we regard the compacted direction as time direction,
the boundary condition for the Polyakov-loop phases can be seen as
imaginary chemical potential.
The periodic and anti-periodic boundary conditions correspond to
different Roberge-Weiss transition points on the QCD phase diagram.
From this viewpoint, it is clear that fundamental fermions with PBC
works to move the vacua to $\mathrm{Re}~\Phi<0$ direction as shown 
in Fig.~\ref{Fig_pl_distribution2} while those with aPBC move it to
$\mathrm{Re}~\Phi>0$ direction as shown in Fig.~\ref{Fig_pl_distribution3}
%%%%%%%%%%%%%%%% Fig %%
%%%%%%%%%%%%%%%%%%%%

\section{Observables comparable to lattice}
\label{sec:OB}

In this section we discuss observables quantitatively in our study,  
which can be compared to existing and on-going lattice simulations.

{\it Mass spectrum} : We fist consider the mass spectrum in the gauge-broken phases, $SU(2)\times U(1)$ and $U(1)\times U(1)$ phases.
As shown in Ref.~\cite{H1,H2,H3,H4}, the Kalza-Klein spectrum for gauge bosons is given by
\begin{equation}
M_{n}^{2}\,=\, {1\over{L^{2}}}(n+q_{i}-q_{j})^{2},
\end{equation}
where $n$ stands for KK index.
We here focus on $n=0$ modes and the case with zero quark mass.
As long as $q_{i}=q_{j}$ $(i\not= j)$, these modes are massless.
On the other hand, when $q_{i}\not =q_{j}$ is realized at the vacuum,
some or all of $n=0$ modes become massive.
This phenomenon is consistent with the Higgs mechanism with
the gauge boson obtaining mass due to gauge symmetry breaking.
In our study, we find the two gauge-broken phase $SU(2)\times U(1)$ and $U(1)\times U(1)$
for $SU(3)$ gauge theory on $R^{3}\times S$.
In the $SU(2)\times U(1)$ phase of $SU(3)$ gauge theory, where we originally have 
8 massless gauge bosons, we have 4 massive modes, whose mass is given by
\begin{equation}
M_{SU(2)\times U(1)}^{2}\,=\,{1\over{4L^{2}}},
\label{su2u1m}
\end{equation}
where we substitute $(q_{1}, q_{2},q_{3})=(0, 0.5,0.5)$ to $q_{i}$.
This result of the gauge boson mass is common for the adj. case and the adj.-fund. case.
Irrespective of the matter field, $SU(2)\times U(1)$ phase has the five gauge bosons with 
the mass (\ref{su2u1m}).
Things change in $U(1)\times U(1)$ phase.
In $U(1)\times U(1)$ phase (re-confined phase) for $(N_{f}, N_{a})=(0,1)$, 
we have 6 massive modes, whose masses are given by
\begin{eqnarray}
M_{U(1)\times U(1)}^{2}(N_f =0, N_a =1)=\left\{ \begin{array}{ll}
&{1\over{9L^{2}}}\,\,\,\,\,\,\,({\rm 4\,\, modes}) \\
& {4\over{9L^{2}}}\,\,\,\,\,\,\,({\rm 2\,\, modes}) \\
\end{array} \right.
\end{eqnarray}
For $(N_{f}, N_{a})=(1,1)$, the $U(1)\times U(1)$ phase (pseudo-confined phase)
again has 6 massive modes, but the masses are different from the above as
\begin{eqnarray}
M_{U(1)\times U(1)}^{2}(N_f =1, N_a =1)=\left\{ \begin{array}{ll}
&{4\over{25L^{2}}}\,\,\,\,\,\,\,({\rm 4\,\, modes}) \\
& {16\over{25L^{2}}}\,\,\,\,\,\,\,({\rm 2\,\, modes}) \\
\end{array} \right.
\end{eqnarray}
Thus, if we look into gauge mass as a function of the compactification scale,
there should be clear difference between different choices of the matter field
even in the same symmetry phase.
It is good indication not only for the gauge symmetry breaking but also 
for specifying the phases.
These results are the case with both four-dimensional and five-dimensional cases.
We note that our results of mass spectrum are valid for small $L$ (weak-coupling),
and the lattice simulation for relatively weak-coupling can reproduce our 
results.

\begin{figure}[htbp]%[H]
\begin{center}
 \includegraphics[width=0.6\textwidth]{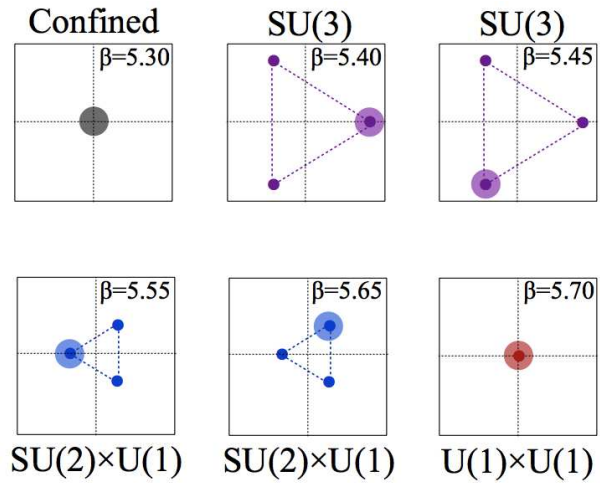}
\end{center}
\caption{Comparison between distribution plot of Polyakov loop $\Phi$ on the lattice \cite{CD1} 
and that of the one-loop effective potential for $SU(3)$ gauge theory on 
$R^{3}\times S^1$ with PBC adjoint quarks. (Watermarks stand for distribution of plot points
on the lattice \cite{CD1}.)
Apart from the strong-coupling confined phase,
all of the specific behavior can be interpreted as the phases we found in our analytical calculations. }
\label{PoD}
\end{figure}

\begin{figure}[htbp]%[H]
\begin{center}
\includegraphics[bb=0 150 850 500, clip, width=0.8\textwidth]{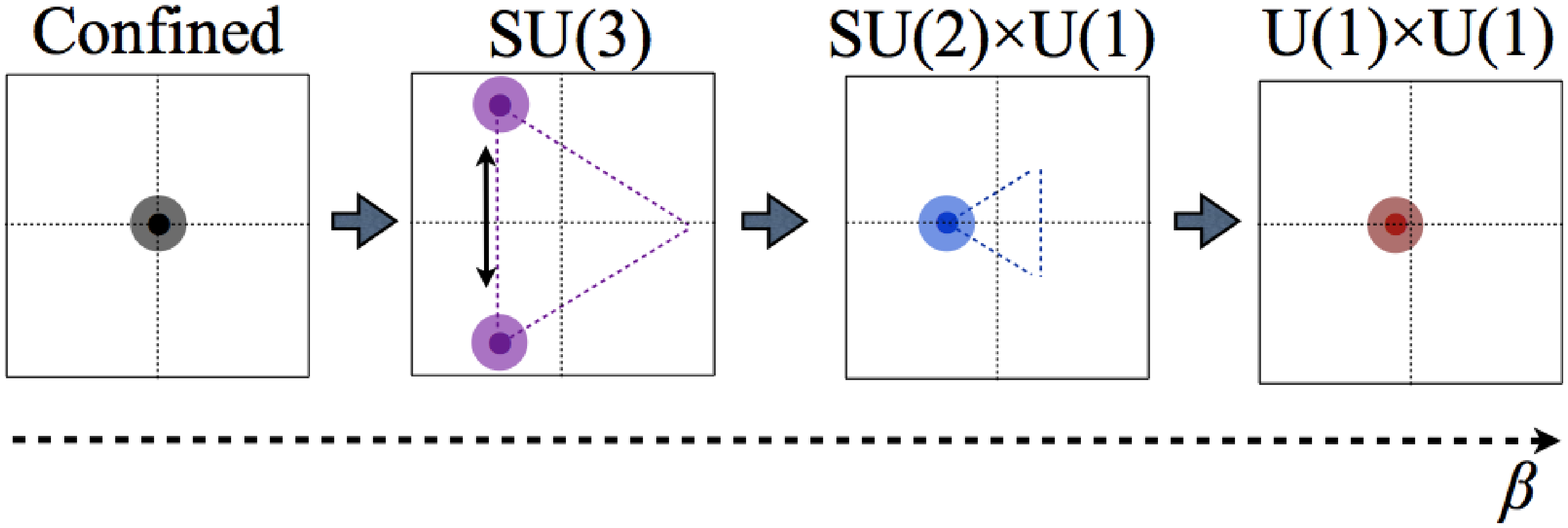} 
\end{center}
\caption{Prediction of distribution plot of Polyakov loop $\Phi$
based on the one-loop effective potential for $SU(3)$ gauge theory on 
$R^{d}\times S^1$ with PBC adjoint and fundamental quarks.}
\label{PoD2}
\end{figure}

{\it Polyakov-loop} : As we have discussed, the trace of compact-dimensional Polyakov loop is also
a good indication of the exotic phases \cite{CD1}.
We first show how the exotic phases seen in the lattice QCD with PBC adjoint fermions \cite{CD1}
is interpreted from the gauge symmetry breaking phases.
Fig.~\ref{PoD} is the distribution plot of the Polyakov loop in the lattice simulations.
Each point corresponds to each of the results for different gauge configurations.
We also depict corresponding results in our analytical calculation of Fig.~\ref{Fig_pl_distribution}
on the figure.
By comparing them,
we find that each of the cases can be understood as one of $SU(3)$ deconfined, $SU(2)\times U(1)$ split or $U(1)\times U(1)$ re-confined phases. (The strong-coupling confined phase cannot
be reproduced in our weak-coupling study.)
The lattice artifacts makes $Z_{3}$ symmetry no-exact, and 
one of $Z_{3}$ minima is selected depending on $\beta$.
This result means that we can predict distribution of Polyakov-loop
for other choices of matter fields in the on-going lattice simulations.
For example, $(N_{f},N_{a})=(1,1)$ case should have the distribution of
Polyakov-loop depending on $\beta$ as shown in Fig.~\ref{PoD2}.
The explicit $Z_{3}$ symmetry breaking shifts the true vacua to the ${\rm Re}~\Phi<0$ side
in the complex space. As a result, the vacuum with ${\rm Re}~\Phi<0$ should be chosen 
in $SU(2)\times U(1)$ phase while the $U(1)\times U(1)$ phase also has ${\rm Re}~\Phi<0$.
This behavior of the Polyakov-loop distribution will be observed in the
on-going simulation for the case with both adj. and fund. quarks.
As shown in Fig.~\ref{Fig_pl_distribution3}, when we take aPBC for fundamental quarks instead of
PBC, the behavior is changed as the vacua at ${\rm Re}~\Phi>0$ are chosen.
These predictions are valid for both four-dimensional and five-dimensional theories.
We note that our results catch physical properties at least at small $L$ (weak-coupling),
and the lattice simulation for relatively weak-coupling can reproduce our results.

%%%%%%%%%%%%%%%% Fig %%
%%%%%%%%%%%%%%%%%%%%

%%%%%%%%%%%%%%%% Fig %%
%%%%%%%%%%%%%%%%%%%%

{\it Chiral condensate and chiral susceptibility} : 
The special behavior of constituent mass (chiral condensate) as a function of $1/L$, 
which we calculated in the PNJL model, is also peculiar to the theories
with spontaneous gauge symmetry breaking.
And it can be detected on the lattice.  
Even if it is not easy to look into exact value and details of constituent mass on the lattice,
the chiral susceptibility works to investigate the subtle behavior.  
Chiral susceptibility in our model is defined as
\begin{equation}
\chi={C_{ll}\over{C_{qq}C_{ll}-C_{ql}^{2}}},
\end{equation}
with
\begin{equation}
C_{qq} = {L^{4}\over{\Lambda}} {\partial^{2}\mathcal{V}\over{\partial m^{2}}},\,\,\,\,\,\,\,
C_{ll} = {L^{4}\over{\Lambda^{3}}} {\partial^{2}\mathcal{V}\over{\partial \Phi^{2}}},\,\,\,\,\,\,\,
C_{qq} = {L^{4}\over{\Lambda^2}} {\partial^{2}\mathcal{V}\over{\partial m \partial \Phi}},
\end{equation} 
where $m$ stands for the adjoint or fundamental quark mass.
For example, the chiral behavior in the $SU(3)$ gauge theory with PBC adjoint matter in Fig.~\ref{Fig_chi_P} indicates that the chiral condensate reacts to the gauge-symmetry phase transitions slightly
and it slowly decreases in the re-confined phase with $1/L$ getting large.
Chiral susceptibility in this case should have two discrete changes at small $1/L$. 
The chiral properties in the $SU(3)$ gauge theory with PBC adjoint and fundamental matters
in Fig.~\ref{Fig_chi_P1} can have more special behavior.
Whichever of the scenarios A and B are realized on the lattice,  peculiar behaviors of
chiral condensate can be detected on the lattice.
We note that our results from the PNJL model are valid for $0\ll 1/L< \Lambda$,
which roughly means the region for $SU(2)\times U(1)$ phase and $U(1)\times U(1)$ phases
at $1/L<\Lambda$ GeV.
We expect the lattice results for these two phases reproduce our results.

%%%%%%%%%%%%%%%Summary%%%%%%%%%%%%%%

\section{Summary}
\label{sec:sum}

In this paper we have studied the phase diagram for $SU(3)$ gauge theories 
with a compact spatial dimension by using the effective models, with emphasis on
the dynamical gauge symmetry breaking.
We show that introduction of adjoint matter with periodic boundary condition leads 
to the unusual phases with spontaneous gauge symmetry breaking, whose ranges are 
controlled by introducing fundamental matter. We also study chiral properties in these theories
and show that the chiral condensate remains nonzero even at a small compacted size.

In Sec.~\ref{sec:EP}, we developed our setup based on one-loop effective potential
and four-point fermion interactions. The effective potential is composed of the gluon, quark
and chiral sectors. The total effective potential corresponds to that of 
the PNJL model with one-loop gluon potential.
This setup is effective enough to investigate vacuum structure and chiral properties 
at weak-coupling or small size of the compact dimension.

In Sec.~\ref{sec:PS4}, we elucidated the vacuum and phase structure
in $SU(3)$ gauge theory on $R^{3}\times S^{1}$.
The theory with PBC adjoint quarks has three different phases in the $L^{-1}$-$m$ space,
including the deconfined phase ($SU(3)$, nonzero $\Phi$), 
the split phase ($SU(2)\times U(1)$, nonzero $\Phi$) and the re-confined phase 
($U(1)\times U(1)$, zero $\Phi$ with nontrivial global minima).
The configuration of these phases in the phase diagram 
is consistent with that of the lattice calculations although we have no confined phase.
By using the PNJL effective potential, we argued that chiral symmetry is slowly restored without
clear phase transition when the size of the compact dimension is decreased. 
In this section we also studied vacuum and phase structure for the case with both 
fundamental and adjoint matters, and showed that the split phase 
is generically widened by adding PBC fundamental quarks. 
We consider that it is because one of the three possible minima for the split phase 
becomes more stable due to the center symmetry breaking.
We showed that another $U(1)\times U(1)$ phase with a negative value of $\Phi$
emerges in this case (pseudo-reconfined phase).

In Sec.~\ref{sec:PS5}, we studied the vacuum and phase structure in 
$SU(3)$ gauge theory on $R^{4}\times S^{1}$.
In the five-dimensional case we concentrate on the one-loop part of the effective potential.
The theory with PBC adjoint quarks again has the split ($SU(2)\times U(1)$) and re-confined 
($U(1)\times U(1)$) phases. Introduction of fundamental quarks works to enlarge the split 
phase more effectively than the four-dimensional case. Especially in the case with one 
adjoint and one fundamental quarks without parity pairs, the split phase overcomes 
the pseudo-reconfined phase and becomes a unique gauge-symmetry-broken phase.

In Sec.~\ref{sec:OB}, we discuss observables comparable to the lattice simulations.
We list up observables including particle mass spectrum, in particular gauge boson mass (Lowest KK spectrum), Polyakov-loop in the compact dimension, constituent mass.
They can be good indications of exotic phase and properties both in the existing and on-going simulations.

All through this paper, we treat PBC and aPBC cases in a parallel manner.
We note that the reference \cite{U1,U2} argues that gauge theory with PBC fermions has no
thermal fluctuation, and thermal interpretation is inappropriate in such a case.
This means that all the results here should be understood as topological phenomena.

In the end of the paper, we discuss whether the lattice simulation can check our predictions. 
One of our main results is that enlargement of the split phase in the presence of
fundamental fermions. This property would be observed in the on-going lattice simulations
on four-dimesional gauge theory with a compact direction \cite{Hgroup1}.
The pseudo-reconfined phase can be also observed where the VEV of Polyakov loop
becomes negative but different from that of the split phase.
On the other hand, it seems difficult to show the first-order phase transition 
between the pseudo-reconfined and split phases.  
The lattice calculation can check whether chiral condensate remains finite
in the re-confined phase even at a very small compacted size.
The small chiral transitions coinciding with the gauge-symmetry
phase transitions are subtle. If the lattice simulation succeeds to measure chiral susceptibility
with high precision, this phenomena may be able to be observed.

During preparing this draft, the collaboration \cite{Hgroup1} kindly informed us 
that it also performed perturbative calculations on the phase diagram in gauge theory
on $R^{3}\times S^{1}$ and $R^{4}\times S^{1}$.
It could be valuable for readers to compare the results from the two independent groups.

\begin{acknowledgments}
The authors are grateful to E.~Itou for giving them chances to have interest 
in the present topics and reading the draft carefully.
They are thankful to Y.~Hosotani, H.~Hatanaka and H.~Kouno 
for their careful reading of this manuscript and valuable comments. 
They also thank G.~Cossu and J.~Noaki for the kind co-operation.
K. K. thanks H. Nishimura for useful comments.
K.K. is supported by RIKEN Special Postdoctoral Researchers Program.
T.M. is supported by Grand-in-Aid for the Japan Society for Promotion of Science (JSPS) 
Postdoctoral Fellows for Research Abroad(No.24-8).
\end{acknowledgments}

\appendix

\section{Vacua and phase structures for other cases}
\label{sec:VSF}

\subsection{Gauge symmetric cases}

\begin{figure}[htbp]%[H]
\begin{center}
 \includegraphics[width=0.4\textwidth]{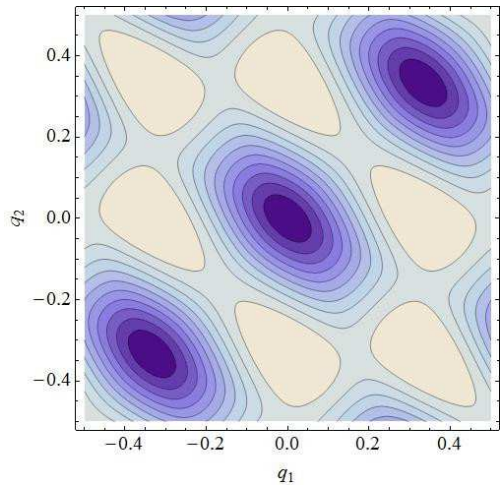}
\end{center}
\caption{
The contour plot of one-loop effective potential for pure $SU(3)$ gauge theory, 
${\cal V}_g L^4$ as a function of $q_1$ and $q_2$.
Thicker region stands for deeper region of the effective potential.
}
\label{Fig_p_g_3D}
\end{figure}

In this appendix, we discuss the vacuum structure
for the several gauge-symmetric cases in $R^{3}\times S^{1}$ $SU(3)$ 
gauge theory. We concentrate on massless cases as $m_a = m_{f} =0$. 
The potential contour plot for $(N_f, N_a)=(0,0)$ (pure gauge theory) 
is shown in Fig.~\ref{Fig_p_g_3D}. 
The effective potential has the minima 
at $(q_1,q_2)=(0,0)$, $(1/3,1/3)$, $(-1/3,-1/3)$.
Since it means that the three Polyakov-line phases are equivalent in the vacuum 
$(q_{1}=q_{2}=q_{3})$, the $SU(3)$ gauge symmetry is intact.

\begin{figure}[htbp]%[H]
\begin{center}
 \includegraphics[width=0.4\textwidth]{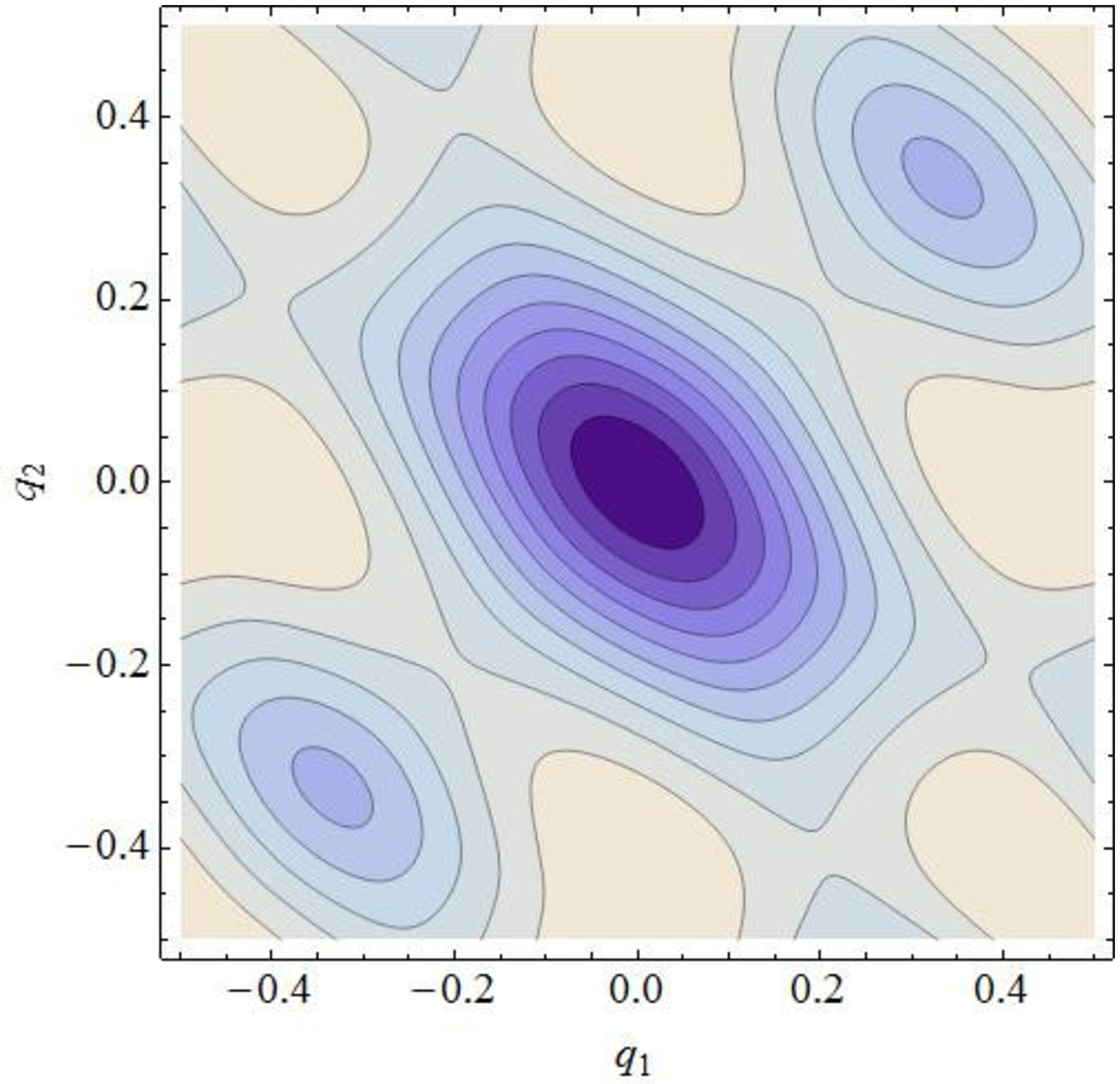}
\end{center}
\caption{
The contour plot of one-loop effective potential for $R^{3}\times S^{1}$ $SU(3)$ gauge theory
with one massless fundamental fermion with anti-periodic boundary condition
($(N_{f},N_{a})=(1,0)$ with aPBC) 
$({\cal V}_g+{\cal V}_f^{1/2}) L^4$.
}
\label{Fig_p_gfa_3D}
\end{figure}

\begin{figure}[htbp]%[H]
\begin{center}
 \includegraphics[width=0.4\textwidth]{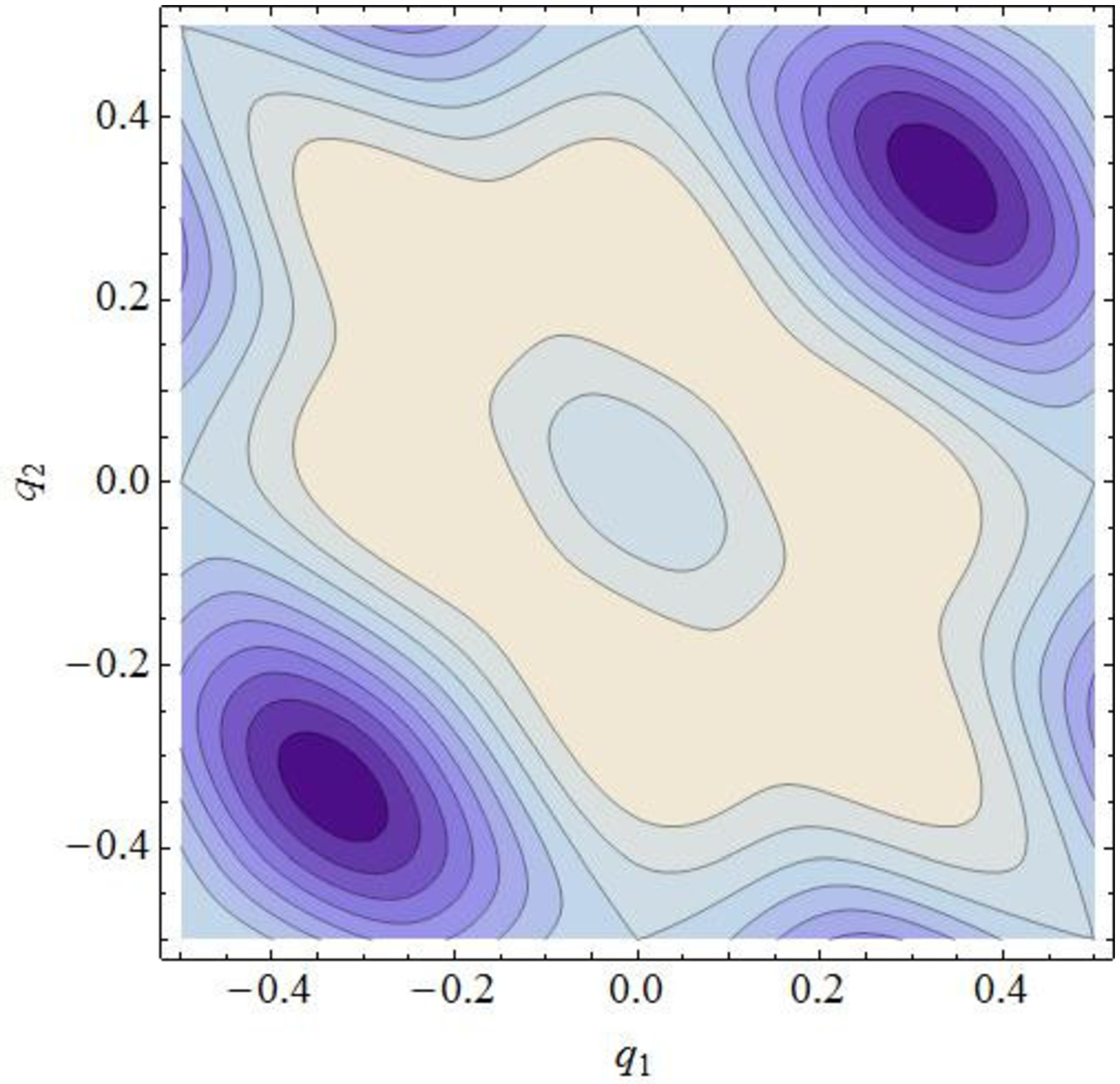}
\end{center}
\caption{
The contour plot of one-loop effective potential for $R^{3}\times S^{1}$ $SU(3)$ gauge theory
with one fundamental fermion with periodic boundary condition
($(N_{f},N_{a})=(1,0)$ with PBC)
$({\cal V}_g+{\cal V}_f^{0}) L^4$.
}
\label{Fig_p_gfp_3D}
\end{figure}

\begin{figure}[htbp]%[H]
\begin{center}
 \includegraphics[width=0.4\textwidth]{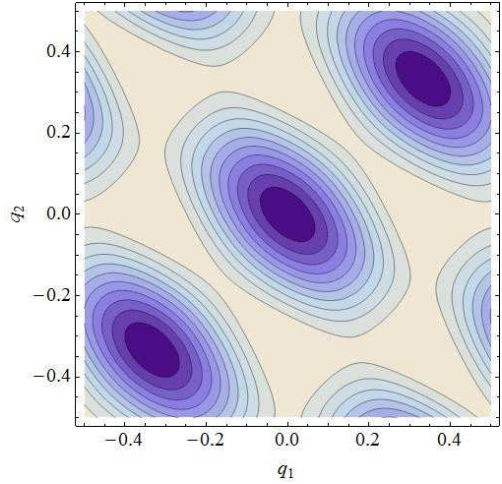}
\end{center}
\caption{
The contour plot of one-loop effective potential for $R^{3}\times S^{1}$ $SU(3)$ gauge theory
with the $N_a = 1$ adjoint fermion with anti-periodic boundary condition
($(N_{f},N_{a})=(0,1)$ with aPBC) $({\cal V}_g+{\cal V}_a^{1/2}) L^4$.
}
\label{Fig_p_gaa_3D}
\end{figure}

The case with one fundamental fermion with anti-periodic
boundary condition is shown in Fig.~\ref{Fig_p_gfa_3D}.
The fundamental quark breaks $Z_3$ symmetry explicitly 
and thus, two of the three minima become the meta-stable.
If the boundary condition of fermion is changed to the periodic one,
the global minima move to $(\pm 1/3, \pm 1/3)$ as shown in
Fig.~\ref{Fig_p_gfp_3D}.
On the other hand, the case for an adjoint fermion with anti-periodic boundary
condition is similar to the pure gauge case as shown in Fig.~\ref{Fig_p_gaa_3D}
since it keeps the $Z_{3}$ center symmetry. 
%%%%%%%%%%%%%%%% Fig %%
%%%%%%%%%%%%%%%%%%%%
%%%%%%%%%%%%%%%% Fig %%
%%%%%%%%%%%%%%%%%%%%
%%%%%%%%%%%%%%%% Fig %%
%%%%%%%%%%%%%%%%%%%%
%%%%%%%%%%%%%%%% Fig %%
%%%%%%%%%%%%%%%%%%%%

\subsection{Phase diagram with non-perturbative deformation}

\begin{figure}[htbp]%[H]
\begin{center}
 \includegraphics[width=0.4\textwidth]{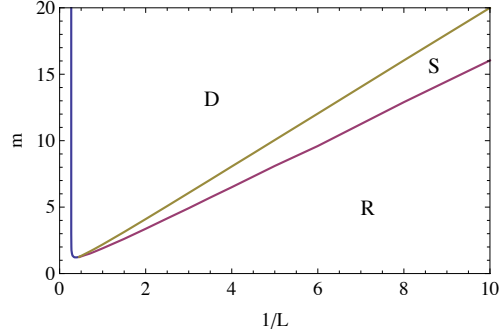}
\end{center}
\caption{Phase diagram for $R^{3}\times S^{1}$ SU(3) theory with one PBC adjoint quark 
based on the effective potential with non-perturbative deformation ($M=596$ MeV).
C stands for ``confined", D stands for ``deconfined" and S for ``split" phases.
Compared to the perturbative one-loop case, the confined phase emerges
but merges into the re-confined phase.}
\label{4d_phase_np}
\end{figure}

In Fig.~\ref{4d_phase_np}, we depict the phase diagram for $R^{3}\times S^{1}$ $SU(3)$ 
gauge theory with $(N_{f}, N_{a})=(0,1)$ with PBC based on 
the nonperturbatively deformed gluonic potential.
As an example, we consider the following modification from the perturbative potential
in \cite{MO3, MeMO1, MeO1, NO1}:
\begin{align}
{\cal V}_{g}^\mathrm{np}
&=- \frac{2}{L^4 \pi^2} \sum_{i,j=1}^N \sum_{n=1}^{\infty} \Bigl( 1 - \frac{1}{N} \delta_{ij} \Bigr)
     \frac{\cos( 2 n \pi q^{ij})}{n^4} \,+\, \frac{M^2}{2\pi^2L^2} \sum_{i,j=1}^N \sum_{n=1}^\infty
  \Bigl( 1 - \frac{1}{N} \delta_{ij} \Bigr) \frac{\cos(2n\pi q^{ij})}{n^2}
\end{align}
where $M$ is the mass-dimension 1 parameter. 
We set the scale parameter as $M=596$ MeV.
The confined phase and the first-order phase transition show up, 
but it is merged into the gauge-broken re-confined phase.
We note that the $SU(3)$ gauge symmetry is explicitly broken in the confined (re-confined) phase. 
%%%%%%%%%%%%%%%% Fig %%
%%%%%%%%%%%%%%%%%%%%

\end{document}